\documentclass[superscriptaddress,showpacs,showkeys,preprintnumbers,nofootinbib,amsmath,amssymb,aps,onecolumn,prd,floatfix]{revtex4-1}

\begin{document}

\title{Superpotential method for chiral cosmological models connected with modified gravity}

\author{S.V.~Chervon}\email{chervon.sergey@gmail.com}
\affiliation{Ulyanovsk State Pedagogical University, Lenin's Square 4/5, Ulyanovsk 432071, Russia}
\affiliation{Bauman Moscow State Technical University, 2nd Baumanskaya Street 5, Moscow 105005, Russia}
\affiliation{Kazan Federal University, Kremlevskaya Street 18, Kazan 420008, Russia}
\author{I.V.~Fomin}\email{ingvor@inbox.ru}
\affiliation{Bauman Moscow State Technical University, 2nd Baumanskaya Street 5, Moscow 105005, Russia}
\author{E.O.~Pozdeeva}\email{pozdeeva@www-hep.sinp.msu.ru}
\affiliation{Skobeltsyn Institute of Nuclear Physics, Lomonosov Moscow State University, Leninskie Gory 1, Moscow 119991,  Russia}
\author{M.~Sami}
\email{msami@jmi.ac.in}  \affiliation{Centre
for Theoretical Physics, Jamia Millia Islamia, New Delhi 110025,
India}\affiliation{Maulana Azad National Urdu
University, Gachibowli, Hyderabad 500032, India}\affiliation{Institute for Advanced Physics $\&$ Mathematics, Zhejiang University of Technology, \\Hangzhou 310032, China}
\author{S.Yu.~Vernov}\email{svernov@theory.sinp.msu.ru}
\affiliation{Skobeltsyn Institute of Nuclear Physics, Lomonosov Moscow State University, Leninskie Gory 1, Moscow 119991,  Russia}

\begin{abstract}
We consider the Chiral Cosmological Models (CCMs) and modified gravity theories associated with them.
Generalization of the superpotential method for a general CCM with several scalar fields is performed, and
the method of construction CCMs admitting exact solutions is developed. New classes of exact
solutions in the two-component CCM connected with an $f(R)$ gravity model with an additional scalar
field have been constructed. We construct new cosmological solutions for a diagonal metric of the
target space, including modified power-law solutions. In particular, we propose the
reconstruction procedure based on the superpotential method and present examples of kinetic part
reconstruction for periodic and hyperbolic Hubble parameters. We also focus on a cyclic type of
Universe dubbed the Quasi-Steady State (QSS) model, with the aim of constructing single- and double-field potentials
for one and the same behaviour of the Hubble parameter using the developed superpotential method for the
CCM. The realization of this task includes a new set of solutions for a CCM with a scale factor
characterized by the QSS theory. We also propose a method for reducing the two-field CCM to the single
scalar field model.
\end{abstract}




\maketitle

\section{Introduction}

Scalar fields play an important role in model building
for early Universe and late time cosmic evolution.
Observations~\cite{cosmo-obser,Bernui:2005pz,Planck2018} show that the Universe evolution can be described by the spatially flat Friedmann--Lema\^{i}tre--Robertson--Walker (FLRW) spacetime as background and cosmological perturbations. Models with scalar fields are well suited  to describe such an evolution. As for
the modified theories of gravity, in general, they can be thought of as the Einstein theory of general relativity plus extra degrees of freedom; for instance, $f(R)$ gravity models correspond to the general relativity models with a single self-interacting scalar field.

Scalar fields are important in inflationary scenarios~\cite{Linde,nonmin,SalopekBond,SBB1989,Lidsey:1995np,inflation2}, including the Starobinsky $R^2$ model~\cite{Starobinsky} and the Higgs-driven inflation~\cite{Higgs}. Models with a single scalar field nonminimally coupled to gravity as well as $f(R)$ gravity models can always be transformed to models with a minimally coupled scalar field with a canonical kinetic term by the metric and scalar field transformations. On the other hand, models with a few fields nonminimally coupled with gravity, in general, do not admit such a transformation~\cite{Kaiser:2010ps}. After the metric transformation, one obtains the Chiral Cosmological Models (CCMs) in the Einstein frame~\cite{Kaiser:2013sna}.  Multifield inflationary models do not contradict  the Planck  data~\cite{Planck2018} and are being actively studied~\cite{Kaiser:2013sna,Gong:2016qmq,KLMattr,MSSM}.
It was recently argued, that compared to single field models, a system with several scalar fields can be better reconciled with observation.
For instance, it was shown in Ref.~\cite{Kaiser:2013sna} that additional degrees of freedom can produce enough power in isocurvature perturbations, which could account for the anomaly in the Planck observation data.
 It is needless to mention that extra degrees of freedom are generic features of   modified theories of gravity.

 At the same time, it has been proven that at least one fundamental scalar field (the Higgs boson) exists. This gives good motivation to consider modified gravity models with an additional scalar field. For example,  inflationary models obtained from $f(R)$ gravity models with additional scalar fields~\cite{Gottlober:1993hp} are very popular now~\cite{delaCruz-Dombriz:2016bjj,Wang:2017fuy,Ema,He:2018gyf,Gorbunov:2018llf}.
The implementation of the inflationary scenario within a well-defined model of particle physics consistent with collider phenomenology would be a fundamental step towards the unification of physics at all energy scales. The Standard model of particle physics that includes only one fundamental scalar field could be an effective limit of some supersymmetry or grant unification theory that has the scalar sector with a few Higgs bosons, and there is no reason to assume that only one scalar field plays a role in the Universe's evolution. In the Einstein frame, all of the  above-mentioned models are CCMs. Quantum motivated higher-order generalizations of general relativity under some conditions can be equivalent to adding several scalar fields to the Einstein-Hilbert action~\cite{Damour1992}.

 The CCM allows one to describe not only the inflationary epoch
of the Universe evolution, but also the present accelerated expansion of
the Universe~\cite{Nojiri:2005pu,Chervon:2013gm,Chervon:2013nsm}.
It has been shown in a recent paper~\cite{Capozziello:2018jya} that the present value of the equation-of-state dark energy parameter has to be phantom-like and for other redshifts, it either has to be a phantom or should have a phantom crossing. It has been shown~\cite{Vikman} that such transitions are physically implausible in one-field models because they are either realized by a discrete set of trajectories in the phase space or are unstable with respect to the cosmological perturbations. This is a strong motivation to consider quintom models~\cite{QuntomREV}, which are a particular case of CCMs.

In the case of single scalar cosmology with a generic potential, the integration procedure is reduced to solving the Ivanov--Salopek--Bond equation~\cite{givanov81,SalopekBond}. There are many methods of obtaining physically relevant solutions, including that with the Higgs potential~\cite{givanov81,Chervon:2017kgn,chervon102,chervon87}.
The reduction of a CCM to a single scalar field model
was proposed in~\cite{chervon26}. A new approach to studying a CCM when gravitational field equations are represented in a linear form under the point transformation was developed in~\cite{Paliathanasis}. For a specific geometry of the target space and a special form of the potential, the way to obtain solutions to the gravitational field equation has been found\footnote{It is difficult to consider such solutions as exact ones without involving the dynamic equations of the chiral fields.}. In the case of several scalar fields with kinetic interaction,  progress in obtaining exact solutions has been achieved for later Universe evolution~\cite{ChervAbb}, the emergent universe~\cite{EmU}, Einstein--Gauss--Bonnet cosmology~\cite{EGB}, and tensor-multiscalar model~\cite{Chervon:2018stfi} as well.

The goal of this paper is to propose a way to get a particular solution of the CCM in the analytic form. We do not seek solutions for a given potential but construct the potential of the scalar field such that the resulting model has exact solutions with important physical properties.
Such a method is similar to the Hamilton--Jacobi method (also known as the superpotential method or the first-order formalism) and is applied to cosmological models with minimally~\cite{givanov81,Muslimov,SalopekBond,Townsend,AKV,Bazeia,AKV2,Andrianov:2007ua,Vernov06,Chervon:2017kgn,Arefeva:2009tkq,Rotova,
Harko:2013gha} and nonminimally~\cite{KTVV2013} coupled scalar fields. This method has been used, in particular, to find exact solutions in single scalar field inflationary models\footnote{Other methods allowing to find exact solutions in inflationary models were presented in~\cite{chervon20E,chervon24,chervon25,chervon26,chervon35E,chervon38E,chervon41E,Nojiri:2005pu,chervon79,chervon87,Barrow:2016qkh,chervon102} (for recent review, see~\cite{Chervon:2017kgn}).}~\cite{Lidsey:1995np,Chervon:2008zz,Binetruy:2014zya,Pieroni:2015cma,Binetruy:2016hna}. Note that a similar method is used for the reconstruction procedure in brane~\cite{DeWolfe,Townsend,MMSV,Bazeia:2013dna,chervon89} and  holographic models~\cite{Gursoy:2008za,Aref'eva:2014sua}.

The key point of the superpotential method is that the Hubble parameter is considered a function of the scalar fields $\phi^A (t)$.
Note that there is an important difference between one-field and multifield models. In the case of one-field models, the above-mentioned procedure is straightforward because only one superpotential (up to a constant) corresponds to the given scalar field $\phi(t)$. In the case of two or more fields, the knowledge of a particular solution $\phi^A (t)$ does not fix the potential. An explicit example of essentially different potentials of two-field models with the same particular solution $\phi^A (t)$ is given in~\cite{AKV2}.
 On the other hand, presenting the Hubble parameter as a function of $K$ scalar fields, which satisfy the first-order equations, one can get a $K$-parametric set of the exact solutions. One parameter corresponds to the shift of time, whereas other parameters correspond to different evolutions of the Universe depending on the initial conditions. An explicit example of a quintom model with a two-parametric set of exact solutions is given in~\cite{Vernov06}.
In this paper, we generalize the superpotential method on the CCM by constructing new models with exact solutions.  In Sections IV-VI two-field CCMs with
two-parametric sets of exact solutions are constructed.

The term {\it  multifield model} is similar to {\it  chiral cosmological model}, which is defined as the self-gravitating nonlinear sigma model with the potential of (self-)interactions employed in cosmology.
Let us mention that the term multiscalar field cosmology was first introduced as the collection of scalar fields with the sum of kinetic (canonical) parts and with the potential depending on all fields. The model with kinetic interaction between the scalar fields is represented in some articles (for example, in the recent work \cite{Paliathanasis}), whereas the term \emph{multifield} was first introduced in the work by V.~de~Alfaro \emph{et al.}~\cite{deAlfaro1979},
with the aim to obtain instanton and meron solutions in a 4D model. They introduced geometrical restriction: all fields take values in the $n$-dimensional sphere. The potential term was not presented in the model, which was called the "four-dimensional sigma model coupled to the metric tensor field".
A.M.~Perelomov in 1981~\cite{Perelomov87} introduced terminology by exchanging the term "group invariant sigma model" for "chiral model", and he also introduced the metric of a chiral model and extended the model from 2D for \emph{N}-dimensional models, so-called chiral models of general type.  In Ref.~\cite{Perelomov87}, it was no connection with gravity. G.G.~Ivanov~\cite{givanov1983}, independently of Ref.~\cite{deAlfaro1979}, came to the "non-linear sigma model coupled to gravity" by considering the Lorentz signature metric of spacetime and scalar (chiral) fields as the source of gravity, besides the kinetic interaction have been introduced as the metric of "chiral" space.

The potential of the interaction of chiral fields was introduced by S.~Chervon in 1994~\cite{Chervon1995}.
Such a model in \cite{Chervon1995} was called the "self-gravitating nonlinear sigma model with the potential". Then in further publications, using terminology introduced by Perelomov,  the model was referred to as the "Chiral inflationary model" and then the "Chiral cosmological model". Thus, the term "Chiral Cosmological Model" reflects the geometrical interactions of fields via the metric of the target (chiral) space which includes the kinetic interactions.

 Let us stress the difference between the pure multifield model (without kinetic coupling and cross interaction between fields)  and the CCM.  Generally speaking, it is impossible to reduce a CCM with a functional component of the target space metric to a conformal Euclidean (or Lorentzian) diagonal metric. For example, in the two-dimensional case, when a surface is embedded in  3D Euclidean space, for a $\textsc{C}^2$-smooth 2D metric component in some neighborhood of the point, it is possible to define the coordinates in which the metric takes the form of conformal Euclidean diagonal metric. But to calculate the form of new coordinates one needs to solve a rather complicated Beltrami equation~\cite{Novikov1990}.

In this paper, we generalize this reconstruction procedure on models with an arbitrary finite number of scalar fields minimally coupled to gravity. The structure of the paper is as follows. In Section~\ref{sec2}, we connect the CCM with modified gravity. In Section~\ref{sp}, the superpotential method develops on CCMs with an arbitrary number of scalar fields. In Sections \ref{SecFR}--\ref{TSRec}, we consider two-component CCMs. In Section~\ref{SecFR}, the considered CCMs correspond to $f(R)$ gravity models with an additional scalar field.
In Sections~\ref{DIAGCTS} and \ref{HCpl1t}, we find models with Ruzmaikin solutions, solutions that correspond to the intermediate inflation and modified power-law solutions for models with the given kinetic terms of the actions due to the choice of the potential.  A procedure for construction of CCMs with trigonometric and hyperbolic Hubble functions due to a suitable choice of the kinetic term is proposed in Section~\ref{TSRec}.
In Section~\ref{QSSCM}, we construct models with the Hubble parameter that describe a cyclic type of Universe dubbed the quasi-steady state.
A method for reducing the two-field CCMs to  single scalar field models is proposed in Section~\ref{2to1}.
Our results are summarized in Section~\ref{sum}.

\section{The connection between chiral cosmological models and the modified gravity}
\label{sec2}
Chiral cosmological models with $K$ scalar fields $\phi^A ~ (\bar{\phi}=\phi^1,\phi^2..., \phi^K) $ are described by the following action,
\begin{equation}
	S=\int d^4x \sqrt{-g}\left[\frac{M^2_{Pl}}{2}R-
		\frac{1}{2}h_{AB}(\bar{\phi})\partial_\mu\phi^A\partial_\nu\phi^Bg^{\mu\nu}
	-V(\bar{\phi})\right], \label{action}
\end{equation}
where the functions $h_{AB}(\bar{\phi})$ and the potential $V(\bar{\phi})$ are differentiable functions,
 $M_{Pl}$ denotes the reduced Planck mass: $M_{Pl}\equiv1/\displaystyle\sqrt{8\pi G}$. We assume that $h_{AB}=h_{BA}$ and that the determinant of this matrix is not equal to zero, so this matrix can be considered
 the field-space metric.

Varying action (\ref{ein-gen}), we get the Einstein equations
\begin{equation}\label{ein-gen}
R_{\mu\nu}-\frac{1}{2}g_{\mu\nu}R=\frac{1}{M_{Pl}^2} T_{\mu\nu},
  \end{equation}
where the energy-momentum tensor is
\begin{equation}\label{emt-gen}
  T_{\mu\nu}=h_{AB}(\bar{\phi})\partial_\mu\phi^A\partial_\nu\phi^B-g_{\mu\nu} \left[ \frac{1}{2}h_{AB}(\bar{\phi})\partial_\rho\phi^A
  \partial_\beta\phi^Bg^{\rho\beta}+V(\bar{\phi}) \right].
  \end{equation}

Variation action \eqref{action} on the chiral field $\phi^C$ leads to the field equation
\begin{equation}
	\frac{1}{\sqrt{-g}}\partial_{\mu}\left( \sqrt{-g}g^{\mu\nu}h_{CB}\phi^B_{,\nu} \right) - \frac{1}{2}g^{\mu\nu} h_{AB,C}\phi^A_{,\mu}\phi^B_{,\nu} - V_{,C}(\bar{\phi}) = 0,
	\label{f-gen}
\end{equation}
where $h_{AB,C}\equiv \frac{\partial h_{AB}}{\partial \phi^C}$.

In the spatially flat FLRW metric with the interval
\begin{equation} \label{FLRW}
	{ds}^{2}={}-{dt}^2+a^2(t)\left(dx_1^2+dx_2^2+dx_3^2\right),
\end{equation}
the Einstein equations (\ref{ein-gen}) have the following form:
\begin{equation}
	\label{a2} 3H^2=\frac{\varrho}{M_{Pl}^2}\,,
\end{equation}
\begin{equation}
	\label{trequ} 2\dot H+3H^2={}-\frac{p}{M_{Pl}^2} \,,
\end{equation}
where
\begin{equation}
	\label{varrho_pressure} \varrho={}-T^0_0=
	\frac{1}{2}h_{AB}(\bar{\phi}){\dot\phi}^A{\dot\phi}^B + V(\bar{\phi}),\qquad
	p=T^1_1=\frac{1}{2}h_{AB}(\bar{\phi}){\dot\phi}^A{\dot\phi}^B - V(\bar{\phi}),
\end{equation}
the dots denote the time derivative and the Hubble parameter  $H(t)$ is the logarithmic derivative of the scale factor: $ H=\dot a/{a}$.

In the FLRW metric \eqref{FLRW}, Eq.~\eqref{f-gen} is transformed into
\begin{equation}
\label{f-1}
	-h_{CB}\left(\ddot{\phi}^B+3H\dot{\phi}^B\right) - h_{CB,D}\dot{\phi}^D\dot{\phi}^B
+ \frac{1}{2}h_{DB,C}\dot{\phi}^D\dot{\phi}^B - V_{,C} = 0.
\end{equation}
Contracting this equation with $h^{AC}$, we obtain\footnote{Note that the upper index \emph{A} could not be moved down with the chiral metric $h_{AB}$.}
\begin{equation}
\label{f-5m}
	\ddot{\phi}^A+3H\dot{\phi}^A +\Gamma^A_{DB}\dot{\phi}^D\dot{\phi}^B + h^{AC}V_{,C} = 0\,,
\end{equation}
where $\Gamma^A_{DB}$ are the Christoffel symbols  for the field-space manifold, defined by the metric~$h_{AB}$.

Let us introduce a new variable
\begin{equation}\label{chi}
X=h_{AB}\dot{\phi}^A\dot{\phi}^B.
\end{equation}

From Eqs.~(\ref{a2}) and (\ref{trequ}), we get
\begin{equation}\label{dHchi}
\dot H={}-\frac{X}{2M_{Pl}^2}\,.
\end{equation}

Multiplying Eqs.~\eqref{f-1} by $\dot\phi^C$, summing them, and using
\begin{equation*}
\dot{X}=2h_{AB}\ddot{\phi}^A\dot{\phi}^B+h_{AB,C}\dot{\phi}^A\dot{\phi}^B\dot{\phi}^C,
\end{equation*}
we get the following equation
\begin{equation}
\label{equchi}
	\frac12\dot{X}+3HX +\dot V= 0.
\end{equation}
Note that Eq.~(\ref{equchi}) is a consequence of Eqs.~(\ref{a2}) and (\ref{dHchi}).

Many modified gravity models are connected with chiral cosmological models. In particular, let us consider models with nonminimally coupled scalar fields, that are described by the following action:
\begin{equation}
S_{J}=\int d^4x\sqrt{-\tilde{g}}\left[f(\bar{\phi})\tilde{R}-\frac{1}{2}\tilde{G}_{AB}\tilde{g}^{\mu\nu}\partial_{\mu}\phi^A\partial_{\nu}\phi^B
-\tilde{V}(\bar{\phi}) \right].
\label{Jordan action}
\end{equation}
By the conformal transformation of the metric
\begin{equation}
g_{\mu\nu}=\frac{2}{M^2_{Pl}}f(\bar{\phi})\tilde{g}_{\mu\nu},
\end{equation}
one gets the following action in the Einstein frame~\cite{SBB1989}:
\begin{equation}
\label{SE}
S_{E}=\int d^4x\sqrt{-g}\left[\frac{M^2_{Pl}}{2}R-\frac12 h_{AB}(\bar{\phi}){g^{\mu\nu}}\partial_\mu\phi^A\partial_\nu\phi^B-V_E\right],
\end{equation}
where
\begin{equation*}
h_{AB}(\bar{\phi})=\frac{M^2_{Pl}}{2f(\bar{\phi})}\left[\tilde{G}_{AB}
+\frac{3f_{,A} f_{,B}}{f(\bar{\phi})}\right],\qquad V_E= M^4_{Pl}\frac{\tilde{V}}{4f^2},
\end{equation*}
and $f_{,A} = \partial f/\partial \phi^A$.

For the class of $f(R)$ gravity models with scalar fields, described by
\begin{equation}
\label{actionFR}
    S_{R} =\int d^4 \tilde{x} \sqrt{-\tilde{g}} \left[f(\bar{\phi}, \tilde{R})-\frac{1}{2}\tilde{g}^{\mu \nu}\tilde{G}_{AB}\partial_{\mu}\phi^A \partial_{\nu}\phi^B -\tilde{V}(\bar{\phi})\right],
\end{equation}
one can introduce an additional scalar field $\phi^{K+1}$ without the kinetic term  and rewrite $S_{R}$ as follows~\cite{Maeda:1988ab}:
\begin{equation}
    \tilde{S}_{J} =\int d^4 \tilde{x} \sqrt{-\tilde{g}} \left[\frac{df(\bar{\phi}, \phi^{K+1})}{d\phi^{K+1}}(\tilde{R}-\phi^{K+1})+ f(\bar{\phi}, \phi^{K+1})-\frac{1}{2}\tilde{g}^{\mu \nu}\tilde{G}_{AB}\partial_{\mu}\phi^A \partial_{\nu}\phi^B -\tilde{V}(\bar{\phi})\right].
\end{equation}

Therefore, we get the model with $K+1$ scalar fields, described by action (\ref{Jordan action}) and can transform it to the chiral cosmological model with action~(\ref{SE}).

\section{The superpotential method for the CCM}
\label{sp}
Generalizing the superpotential method for multifield models with the standard kinetic term~\cite{SalopekBond} and for two-field models with a constant kinetic term~\cite{AKV2,Vernov06,Arefeva:2009tkq}, we describe the superpotential for the general CCM.

We assume that  functions $\phi^A$ are solutions of the
following system of $K$ ordinary differential equations:
\begin{equation}
	\label{FG}
\frac{d \phi^A}{dt}\equiv \dot\phi^A=F^A(\bar\phi).
\end{equation}

In this case, the Hubble parameter $H(t)$ is a function of all scalar fields,
\begin{equation*}
H(t) = W(\bar{\phi})+C_W,
\end{equation*}
where superpotential $W$ is a differentiable function and $C_W$ is a constant part of the Hubble function that plays a special role (see, for example, \cite{Chervon:2017kgn}). It is convenient to write $C_W$  separately.
Thus, Eqs.~\eqref{a2} and \eqref{trequ} take the following form:
\begin{equation}
	3M^2_{Pl}\left[W(\bar\phi) +C_W\right]^2 = \frac{1}{2}h_{AB}F^AF^B + V(\bar{\phi}),
\label{e5-2}
\end{equation}
\begin{equation}
	M^2_{Pl}\left(2W_{,A}F^A + 3\left[W(\bar\phi) +C_W\right]^2\right) ={}-\frac{1}{2}h_{AB}F^AF^B + V(\bar{\phi}).
\label{e6}
\end{equation}

Subtracting Eq.~(\ref{e5-2}) from Eq.~(\ref{e6}), we get
\begin{equation}
	\left(W_{,A}{}+\frac{1}{2M^2_{Pl}}h_{AB}F^B \right) F^A=0\,.	\label{e6a2}
\end{equation}

One can see that the sufficient conditions to satisfy Eq.~(\ref{e6a2}) are the following relations:
\begin{equation}
\label{WF2}
    W_{,A}={}-\frac{h_{AB}F^B }{2M^2_{Pl}}\,,
\end{equation}
for all $A$.
This is rather tight restriction which is equivalent to the decomposition method used in the
works~\cite{Chervon:2017kgn, ChervAbb,Chervon:2018stfi}.

Using
$\ddot\phi^B=\dot{F}^B=F^B_{,D}F^D$,
 we rewrite the field equations~\eqref{f-1} as follows:
\begin{equation}
	h_{CB}F^B_{,D}F^D + h_{CB,D}F^DF^B - \frac{1}{2}h_{DB,C}F^DF^B +
3[W(\bar\phi) +C_W]h_{CB}F^B + V_{,C} = 0\,.
	\label{f-5}
\end{equation}

From Eq.~\eqref{e5-2}, it  follows that
\begin{equation}\label{V,C}
V_{,C}= 6M^2_{Pl}[W(\bar\phi)+C_W]W_{,C}-\frac12h_{DB,C}F^DF^B-h_{DB}F^D_{,C}F^B.
\end{equation}

Substituting this expression of $V_{,C}$ into Eq.~\eqref{f-5}, we get
\begin{equation}\label{W-field}
3[W(\bar\phi) +C_W]\left\{2M^2_{Pl}W_{,C}+h_{DC}F^D \right\}
=(h_{DB}F^D_{,C}-h_{DC}F^D_{,B})F^B+(h_{DB,C}-h_{CB,D})F^DF^B.
\end{equation}

If condition~\eqref{WF2} is satisfied, then
\begin{equation}
\label{cond1}
\left[h_{DB}F^D_{,C}-h_{DC}F^D_{,B}\right]F^B+(h_{DB,C}-h_{CB,D})F^DF^B=0.
\end{equation}

Also, from the obvious equality $W_{,CB}=W_{,BC}$ we get
\begin{equation}
\label{cond2a}
    (h_{BD,C}-h_{CD,B})F^D=h_{CD}F^D_{,B}-h_{BD}F^D_{,C}.
\end{equation}
Condition (\ref{cond1}) is a consequence of (\ref{cond2a}). The matrix $h_{AB}$ is symmetric and the conditions (\ref{cond2a}) are trivial at $B=C$, so it is enough to check Eq.~(\ref{cond2a}) for all $B<C$ only.

Thus, the task of solving the dynamic equations of the model is reduced to  Eqs.~\eqref{e5-2}, \eqref{WF2}, and \eqref{cond2a}. The last equation guarantees that the solution of chiral field equation~\eqref{f-5} is true.

Further we will use the expression for the potential $V(\bar{\phi})$ in terms of the superpotential
$W(\bar{\phi})$, which follows from Eqs.~(\ref{e5-2}) and (\ref{WF2}):
\begin{equation}
V(\bar{\phi})=3M^2_{Pl}[W(\bar\phi) +C_W]^2+M^2_{Pl}W_{,A}F^A\,.
\label{V-W}
\end{equation}

Applying Eq.~\eqref{WF2} once more, the physical potential can be presented in the form
\begin{equation}
V(\bar{\phi})=3M^2_{Pl}[W(\bar\phi) +C_W]^2-2M^4_{Pl}h^{AB}W_{,A}W_{,B}\,.
\label{V-W-2}
\end{equation}

From Eq.~\eqref{f-5m}, we obtain
\begin{equation}
	\dot{F}^E+ \Gamma^E_{DB} F^DF^B + 3WF^E + h^{CE}V_{,C}=0\,.
	\label{f-7}
\end{equation}

To demonstrate how one can  get exact solutions due to the superpotential method we consider a  diagonal matrix $h_{AB}$ such that
each $h_{BB}$ depends only on $\phi^A$, with $A\leqslant B$, and has the following form:
\begin{equation}\label{h11diag}
h_{11}=s_1(\phi^1),\qquad h_{BB}=u_B(\phi^1,\dots,\phi^{B-1})s_B(\phi^B),
\end{equation}
for all $B=2,\dots,K$.

Let us prove,  for any $h_{BB}$ given by (\ref{h11diag}), that we can construct the CCM with exact solutions obtained either in analytic form or in quadratures. We choose
\begin{equation}
\label{Wsum}
    H(t)=\sum_{A=1}^K W_A(\phi^A)+C_W,
\end{equation}
where $W_A(\phi^A)$ are differentiable functions.

The function $W$ fixes the potential $V$ by Eq.~(\ref{V-W}).
For any $A$, the function $F^A$ is defined as follows:
\begin{equation}
\label{FA}
F^A={}-2M^2_{Pl}\frac{W_{A,A}}{h_{AA}}.
\end{equation}
Let us check  conditions~(\ref{cond2a}).
For the matrix $h_{AB}$, defined by (\ref{h11diag}) and $B<C$, we get
\begin{equation}
   \frac{d}{d\phi^B}\left(h_{CC}F^C\right)=0,
\end{equation}
without summing on $C$. It is easy to see that the functions $F^A$, defined (for each index A) by (\ref{FA}) satisfy these conditions.
Therefore, we obtain all functions $F^A$, and system (\ref{FG}) takes the following form:
\begin{equation}
\label{Fsyst}
\begin{split}
\dot\phi^1&={}-2M^2_{Pl}\frac{W_{1,1}(\phi^1)}{h_{11}(\phi^1)},\\
\dot\phi^2&={}-2M^2_{Pl}\frac{W_{2,2}(\phi^2)}{h_{22}(\phi^1,\phi^2)},\\
\dots &\\
\dot\phi^K&={}-2M^2_{Pl}\frac{W_{K,K}(\phi^K)}{h_{KK}(\phi^1,\phi^2,\dots,\phi^K)}.
\end{split}
\end{equation}

This system can be solved at least in quadrature. Indeed, the first equation of this system, as the first order autonomous differential equation, can be solved in quadrature. Let us assume that all $\phi^A$ are known for $A<B$ and consider the equation for $\phi^B$. Using $h_{BB}=u_B(\phi^1,\dots,\phi^{B-1})s_B(\phi^B)$, we get
\begin{equation}
  \frac{s_B(\phi^B)}{W_{B,B}(\phi^B)}d\phi^B= {}-\frac{2M^2_{Pl}}{u_B(\phi^1(t),\dots,\phi^{B-1}(t))}dt.
\end{equation}
This equation is integrable for any $u_B$ given as a function of $t$. So, the solution is found, and the statement is proven by induction.

Note that the superpotential method allows us to construct such a one-parametric set of the CCM that the corresponding Hubble parameters differ on an arbitrary constant $C_W$. We have shown that any of these CCMs has  a $K$-parametric set of exact solutions that can be found explicitly or in quadrature.

In the next sections, we restrict ourselves to two-dimensional $h_{AB}$. In particular, we show that two-dimensional $h_{AB}$ in the form (\ref{h11diag}) naturally arise from $f(R)$ gravity models with one scalar field.

\section{Exact solutions for an $f(R)$ gravity model with an additional scalar field}
\label{SecFR}

In this and the following sections, we consider two-component CCMs and
denote
\begin{equation*}
\phi^1=\psi, \quad \phi^2=\chi, \quad  \dot{\phi}^1=F^1=U(\psi,\chi),\quad  \dot{\phi}^2=F^2=S(\psi,\chi).
\end{equation*}

It was shown in Refs.~\cite{He:2018gyf,Gorbunov:2018llf} that under the metric transformation
\begin{equation*}
    g_{\mu \nu}=e^{\frac{\sqrt{2}\psi}{\sqrt{3}M_{Pl}}} \tilde{g}_{\mu \nu},
\end{equation*}
 the $f(R)$ gravity model with a scalar field $\chi$, described by the action
\begin{equation}
\label{actionR2}
    S_{\text{J}} =\int d^4 \tilde{x} \sqrt{-\tilde{g}} \left[f(\chi, \tilde{R})-\frac{1}{2}\tilde{g}^{\mu \nu}\partial_{\mu}\chi \partial_{\nu}\chi \right],
\end{equation}
transforms to the chiral cosmological model, described by action~(\ref{action}) with
\begin{equation}
		h_{AB}=
	\begin{pmatrix}
	1 & 0 \\
	0 & K(\psi) \\
	\end{pmatrix}
\end{equation}
where
\begin{equation}
\label{fRscalar}
     \psi=\sqrt{\frac{3M^2_{Pl}}{2}} \ln \left(\frac{2}{M^2_{Pl}}\left|\frac{\partial f}{\partial \tilde{R}}\right|\right), \qquad K(\psi)=e^{C\psi},
\end{equation}
and there is a constant $C={}-\sqrt{\frac{2}{3M^2_{Pl}}}$.

Let us consider this CCM to obtain exact solutions due to the superpotential method described in the previous section.

From Eq.~(\ref{WF2}), we get
\begin{equation}
    W_{,1}\equiv W_{,\psi}={}-\frac{1}{2M^2_{Pl}}U(\psi,\chi),\qquad W_{,2}\equiv W_{,\chi}={}-\frac{K(\psi)}{2M^2_{Pl}}S(\psi,\chi).
\end{equation}

To get $W$ in the form (\ref{Wsum}) we choose
\begin{equation}
  U(\psi,\chi)=U(\psi),\qquad  S(\psi,\chi)=\frac{Q(\chi)}{K(\psi)},
\end{equation}
where $Q(\chi)$ is an arbitrary function.

The superpotential $W$ is defined as follows:
\begin{equation}
 W={}-\frac{1}{2M^2_{Pl}}\left(\int U d\psi+\int Q d\chi\right).
\end{equation}

To get an exact particular solution we assume the explicit form of the functions $U(\psi)$ and $Q(\chi)$ and solve the corresponding system of the first-order differential equations:
\begin{equation}
    \dot{\psi}=U(\psi),\qquad \dot{\chi}=\frac{Q(\chi)}{K(\psi)}.
\end{equation}

We choose
\begin{equation}
    U(\psi)=F_0 \mathrm{e}^{-\Lambda_1\psi}, ~~~
\end{equation}
where $F_0$ and $\Lambda_1$ are constants. Therefore,
\begin{equation}
\psi(t)=\frac{1}{\Lambda_1}\ln(\Lambda_1 F_0 (t-t_0))
\end{equation}
and
\begin{equation}
K(\psi)=e^{C\psi(t)}=(\Lambda_1F_0(t-t_0))^{C/\Lambda_1}.
\end{equation}

Substituting the obtained $\psi(t)$, we get $\chi(t)$.
For example, for
\begin{equation}\label{FQ}
Q(\chi)=F_0e^{-\Lambda_2\chi}, ~~~ \Lambda_2=\mathrm{const},
\end{equation}
we obtain
\begin{equation}
    \chi(t)={}-\frac{1}{\Lambda_2}\ln\left(\frac{C-\Lambda_1}{\Lambda_2(C_2-[\Lambda_1F_0(t-t_0)]^{1-C/\Lambda_1})}\right),
\end{equation}
where $C_2$ is an integration constant.

The Hubble parameter
\begin{equation}
H(t)=W_1(\psi)+W_2(\chi)+C_W,
\end{equation}
where
\begin{equation}\label{W1W2}
    W_1(\psi)=\frac{F_0}{2\Lambda_1M^2_{Pl}}e^{-\Lambda_1\psi},\qquad
    W_2(\chi)=\frac{F_0}{2\Lambda_2M^2_{Pl}}e^{-\Lambda_2\chi}.
\end{equation}

So, the considered model with the potential
\begin{equation}\label{Vphi1phi2}
    V=\frac{3}{4M^2_{Pl}\Lambda_1^2\Lambda_2^2}\left(\Lambda_2F_0 e^{-\Lambda_1\psi}+\Lambda_1F_0e^{-\Lambda_2\chi}+2M_{Pl}^2\Lambda_1\Lambda_2C_W\right)^2-\frac{F_0^2}{2}\left( e^{-C\psi-2\Lambda_2\chi}+e^{-2\Lambda_1\psi}\right),
\end{equation}
 has exact solutions with the following Hubble parameter:
\begin{equation}
H= C_W+\frac{1}{2\Lambda_1^2M^2_{Pl}(t-t_0)}+\frac{F_0(C-\Lambda_1)}{2(\Lambda_2^2M^2_{Pl}\left[C_2-(\Lambda_1F_0(t-t_0))^{1-C\Lambda_1}\right])}.
\end{equation}
Note that the Hubble parameter depends on two integration constants, $t_0$ and $C_2$.
We get a one-parametric set of models with exact solutions.

Let us check that for some values of model parameters we get a slow-roll regime that maybe suitable for inflation.
We assume that $\Lambda_1>0$ and $\Lambda_2>0$. In this case, the potential has a finite non-negative limit
\begin{equation*}
    V\rightarrow {3M^2_{Pl}C_W^2}
\end{equation*}
at the scalar fields, which tend to plus infinity such that $2\Lambda_2\chi>-C\psi$.

We consider large initial values of the scalar fields and assume that the scalar fields monotonically decrease during inflation,  choosing $F_0<0$.
For large positive values of the scalar fields we have quasi de Sitter solutions with $H\simeq  C_W$. So, we get a model that looks suitable for describing inflation. Full analysis of the possible inflationary scenarios with calculations of the inflationary parameters will be a subject of further investigation.

\section{Diagonal constant metric of the target space}
\label{DIAGCTS}

Studying a canonical scalar field equation, we set the metric coefficient $h_{11}$ equal to unity: $ h_{11}=1$.
Therefore, it will be of interest to study the diagonal metric with a constant chiral metric component, $h_{22}=1$.

Equation \eqref{e6a2} takes the form
\begin{equation}\label{EdH}
  \frac{\partial W}{\partial \psi}U(\psi,\chi)+\frac{\partial W}{\partial \chi}S(\psi,\chi)=-\frac{1}{2}U^2-\frac{1}{2}S^2.
\end{equation}

Let us insert the metric components of the target space into the field equation \eqref{W-field}. After simple algebra, we obtain
\begin{equation}\label{f-c1}
3W\left( 2W_{,\psi}+U(\psi,\chi)\right)=S_{,\psi}S-U_{,\chi}S,
\end{equation}
\begin{equation}\label{f-c2}
3W\left( 2W_{,\chi}+S(\psi,\chi)\right)=U_{,\chi}U-S_{,\psi}U.
\end{equation}

If we suggest that $U=U(\psi)$ and $S=S(\chi)$, then  Eqs.~\eqref{f-c1} and \eqref{f-c2} are reduced to
\begin{equation}\label{f-c1red}
   2W_{,\psi}+U(\psi)=0\,,
\end{equation}
\begin{equation}\label{f-c2red}
  2W_{,\chi}+S(\chi)=0\,.
\end{equation}

Let us note that the consistency relation $W_{,\psi \chi }=W_{,\chi\psi} $ is satisfied.

Such a representation gives us the possibility to perform integration and find the superpotential $W(\psi,\chi)$:
\begin{equation}
    W={}-\frac{1}{2M^2_{Pl}}\left(\int U(\psi) d\psi+
    \int S(\chi)d\chi\right).
\end{equation}

 Let us choose the linear dependence of the chiral field derivatives
\begin{equation}\label{1-1-ft}
U(\psi)=\mu_1 \psi+c_1,~~~S(\chi)=\mu_2 \chi+c_2\,.
\end{equation}

From here one can find the chiral field evolution
\begin{eqnarray}
  \psi &=& \frac{1}{\mu_1}e^{\mu_1 t}-\frac{c_1}{\mu_1}, \\
  \chi &=&  \frac{1}{\mu_2}e^{\mu_2 t}-\frac{c_2}{\mu_2}.
\end{eqnarray}

Then the superpotential can be obtained by the integration of \eqref{WF2} and reads
\begin{equation}\label{1-1-W}
W(\psi, \chi)={}-\frac{1}{2}\left( \frac{\mu_1}{2} \psi^2 +c_1\psi+\frac{\mu_2}{2} \chi^2 +c_2 \chi \right).
\end{equation}

Further inserting the chiral fields into Eq.~\eqref{1-1-W}, we can obtain the Hubble function $H=H(t)$  in two-parametric form:
\begin{equation}\label{1-1-H}
  H(t)= {}-\frac{1}{4}\left( \frac{e^{2\mu_1 t}}{\mu_1}- \frac{c_1^2}{\mu_1}+\frac{e^{2\mu_2 t}}{\mu_2}-\frac{c_2^2}{\mu_2}\right).
\end{equation}
Thus, we obtain a double exponent solution for the scalar factor.

Let us choose the following dependence:
\begin{equation}\label{k1-k2}
  U(\psi)=\mu_1 \psi^{k_1},\qquad S(\chi)=\mu_2 \chi^{k_2}.
\end{equation}

The exceptional situation $k_1=k_2=0$ leads to a very interesting solution,
\begin{equation}\label{1-1-W1}
W(\psi, \chi)={}-\frac{1}{2}\left( \mu_1^2+\mu_2^2 \right)t.
\end{equation}
Remembering that $H(t)=W(\psi,\chi)+C_W$, we find
\begin{equation}\label{k-0}
  a(t)=a_*\exp \left[ -\frac{1}{4}\left( \mu_1^2 +\mu_2^2 \right)t^2+C_Wt \right]
\end{equation}
This scale factor corresponds to Ruzmaikin's solutions~\cite{Ruzmaikina,Skugoreva:2014gka}.

When $ k_1,\, k_2\neq 0,1$, we have solutions for the fields
\begin{equation}\label{k1-neq}
\psi (t)=\left[ (1-k_1)(\mu_1 t+c_1)\right]^{\frac{1}{1-k_1}},
\end{equation}
\begin{equation}\label{k2-neq}
\chi (t)=\left[ (1-k_2)(\mu_2t+c_2)\right]^{\frac{1}{1-k_2}}.
\end{equation}

The Hubble function,  once again in two-parametric form, is
\begin{equation}\label{H-k-neq}
 H (t)={}-\frac{1}{2}\left[ \frac{\mu_1}{k_1+1}\left[(1-k_1)(\mu_1t+c_1)\right]^{\frac{k_1+1}{1-k_1}} +  \frac{\mu_2}{k_2+1}\left[(1-k_2)(\mu_2t+c_2)\right]^{\frac{k_2+1}{1-k_2}} +C_W \right].
\end{equation}

The scale factor is
\begin{equation}\label{a-k-neq}
  a(t)=a_*\exp \left[\frac{\mu_1}{k_1+1}(1-k_1)^{\frac{k_1+1}{1-k_1}}\left[(\mu_1t+c_1)\right]^{\frac{2}{1-k_1}} +  \frac{\mu_2}{k_2+1} (1-k_2)^{\frac{k_2+1}{1-k_2}} \left[(\mu_2t+c_2)\right]^{\frac{2}{1-k_2}} +C_W t\right].
\end{equation}

Thus, the obtained solution corresponds to the intermediate inflation.

\section{Modified power-law solutions}
\label{HCpl1t}
As known, solutions with the power-law Hubble parameter correspond to the radiation and matter dominated epochs.
It is interesting to get exact solutions with the Hubble parameter $H=C_0+C_1/t$, where constants $C_i$ can be chosen in such a way that
the solution has both dark matter and dark energy parts.

We consider the CCM with the metric of the target space
\begin{eqnarray}
h_{AB}=\left(
\begin {array}{cc} {\frac {{C_1}M^{2}_{Pl}}{\psi^{2}}}&0
\\ \noalign{\medskip}0&{\frac {{C_2}M^{2}_{Pl}}{\chi^{2}}}
\end {array} \right),
\end{eqnarray}
and assume the following form of the superpotential
\begin{equation}
W={}-{Y_1}\psi^{{m_1}}-{Y_2}\chi^{{m_2}},
\end{equation}
where $Y_i$ and $m_i$ are constants.

Using expression \eqref{V-W}, we get  the  potential
\begin{eqnarray}
V&=&3M^2_{Pl}C_W^{2}-6M^2_{Pl}{C_W}{Y_1}\psi^{m_1}-6M^2_{Pl}{C_W}{Y_2}\chi^{{m_2}}+6\,
M^2_{Pl}{Y_1}\psi^{{m_1}}{Y_2}\chi^{{m_2}}\nonumber\\&+&{\frac {M^2_{Pl}Y_1^{2}\left(3C_1 -2m_1^{2} \right)   }{{C_1}}}\psi^{2{m_1}}+\frac{M^2_{Pl}Y_2^{2}\left(3{C_2} -2m_2^{2} \right)   }{C_2}\chi^{2{m_2}}\,.
\end{eqnarray}

Equations \eqref{WF2} are:
\begin{equation}
\dot{\psi}=2\,{\frac {{Y_1}{m_1}}{{C_1}}}\psi^{{m_1}+1},\qquad \dot{\chi}=2\,{\frac {{Y_2}{m_2}}{{C_2}}}\chi^{{m_2}+1}.
 \end{equation}
This system has the following solution:
\begin{equation}
\label{phichisol}
\psi= \left( \frac{{}-C_1} {2Y_1\,m_1^{2}(t-t_0)} \right)^{1/{{m_1}}}
 ,\qquad  \chi=\left( {\frac{{}-{C_2}}{2{Y_2}m_2^{2}(t-\tilde{t}_0)}}\right)^{1/m_2},
\end{equation}
where $t_0$ and $\tilde{t}_0$ are integration constants.
Substituting the obtained solution for the superpotential, we get the Hubble parameter
\begin{equation}
\label{TPH}
H(t)=C_W+\frac{{C_1}}{2m_1^{2}(t-t_0)}+\frac{{C_2}}{2m_2^{2}(t-\tilde{t}_0)}.
\end{equation}

Choosing $t_0=0$ and $\tilde{t}_0=0$, we get
\begin{equation}
H={C_W}+\frac{1}{2t}\left(\frac{C_1}{m_1^2}+\frac{C_2}{m_2^2}\right).
\end{equation}

So, the Hubble parameter is a sum of a constant that corresponds to the dark energy dominant epoch and the power-law function
that corresponds to the radiation dominant epoch at $\frac{C_1}{m_1^2}+\frac{C_2}{m_2^2}=1$ or to the matter dominant epoch at
 $\frac{C_1}{m_1^2}+\frac{C_2}{m_2^2}=\frac43$. Choosing $\tilde{t}_0\neq t_0$, we get more complicated solutions.  So, starting from an exact solution with  $H=C_0+C_1/t$ and using the superpotential method, we not only reconstruct the corresponding potential, but also find that the model obtained has a two-parametric set of the exact solutions, described by formulae~(\ref{phichisol}) and (\ref{TPH}).

\section{Target space reconstruction from the superpotential}
\label{TSRec}

\subsection{The search of the CCM with the given $H(t)$}

From the given superpotential (Hubble function) it is possible to reconstruct the kinetic part to get the exact solution of the model.

Considering single scalar field cosmology, we may find a different formulation of the problem because we have two independent differential equations with three functions. Therefore, we must choose which function we have to consider as the given one. To fix the potential energy (or, simply, the potential) is preferable because it may be collected from  high energy physics.
If we study a canonical scalar field, we suggest the unit multiplier in the kinetic energy term. Extending such an approach to CCMs, we assume that the metric of a chiral space should be fixed and the analogue to the canonical field will be the unit diagonal metric. In the models with symmetry (for example, $SO(3)$---invariant CCM) the chiral metric is fixed also.

Another approach can be proposed when we use the superpotential method. If we are looking for the CCM which obeys the given superpotential (or, equivalently, the Hubble function)  we can use the so-called deformation of a chiral space method. That is, we define such metric components which  match the given data. Such an approach is similar, in some sense, to "the fine tuning of the potential" method for a single scalar field, where a given Hubble function allows one to define the potential and kinetic energies.
Analyzing Eq.~\eqref{WF2}, which takes for the first field $\psi$ the following form
\begin{equation}\label{WF2-1}
 2 M_{Pl}^2W_{,\psi} = {}-h_{11}(\psi)\dot{\psi},
\end{equation}
one can come to the conclusion that the functional part of the lhs may be included in the chiral metric component $h_{11}(\psi)$ and one can set $ \dot{\psi} $ equal to unity: $\dot{\psi} =1$. This gives us the possibility (performing the same procedure for the second field $\chi$)
to choose the linear dependence of the fields on time:
\begin{equation}\label{rec-2}
  \psi= t+\psi_*,\qquad\chi=t+\chi_*.
\end{equation}

\subsection{Examples of periodic Hubble functions}
To demonstrate such an approach let us study an example of the following periodic Hubble function
\begin{equation}
\label{Hsin}
H(t)=H_0 \sin (\lambda t)+C_W.
\end{equation}

We may represent the superpotential in the following form,
\begin{equation}\label{rec-1}
W(\psi,\chi)=H_0\left[ (1-\lambda_0)\sin(\lambda \psi) +
\lambda_0 \sin (\lambda\chi)\right],
\end{equation}
where $H_0$, $\lambda$, and $\lambda_0$ are constants.

The field equations  \eqref{WF2}  are nothing but the definition of the chiral metric component
\begin{equation}
\label{rec-4}
\begin{split}
 h_{11}(\psi)&={}-2H_0M_{Pl}^2 (1-\lambda_0)\lambda\cos(\lambda\psi),\\
h_{22}(\chi)&={}- 2H_0 M_{Pl}^2\lambda_0\lambda\cos(\lambda\chi).
\end{split}
\end{equation}

 The potential can be defined by \eqref{V-W-2}:
 \begin{equation}
 \label{rec-5}
 \begin{split}
 V(\psi,\chi)&= 3H_0^2 M_{Pl}^2\left[ (1-\lambda_0)\sin (\lambda \psi)+\lambda_0\sin(\lambda \chi)+\frac{C_W}{H_0}\right]\\
  &{}+H_0 M_{Pl}^2\lambda\left( (1-\lambda_0)\cos(\lambda\psi)+\lambda_0\cos(\lambda\chi) \right)
 \end{split}
 \end{equation}

It is interesting to note that  solution \eqref{rec-4} belongs to the two-field case in \eqref{Fsyst} with $u=1$. Putting $\psi_*=\chi_*=0$,
 we obtain the Hubble parameter (\ref{Hsin}). If $\psi_*=\chi_*=-\pi/(2\lambda)$, then we get
\begin{equation}
\label{Hcos}
H(t)=H_0 \cos (\lambda t)+C_W.
\end{equation}
The model constructed has a two-parametric set of exact solutions with
 \begin{equation}
\label{Hsingen}
H(t)=H_0\left[ (1-\lambda_0)\sin(\lambda (t+\psi_*)) +
\lambda_0 \sin (\lambda(t+\chi_*))\right]+C_W.
\end{equation}

Note that the case
\begin{equation}
H(t)=H_0 \sin^2 (\lambda t)+C_W
\end{equation}
is reduced to the previous one due to the substitution $\sin^2(\lambda t)=1-2\cos(2\lambda t)$.

For
\begin{equation}\label{Hexpsin}
H(t)=H_0 \exp (-\alpha t \sin t)
\end{equation}
the same approach gives us the solution
\begin{eqnarray}
 H_0^{-1} W(\psi,\chi) &=& (1-\lambda_0) \exp (-\alpha \psi \sin \psi)+\lambda_0 \exp (-\alpha \chi \sin \chi),\\
  h_{11}(\psi) &=&2H_0M_{Pl}^2\alpha (1-\lambda_0)\left[ \sin(\psi)+\psi\cos(\psi) \right]\exp (-\alpha \psi \sin(\psi)),\\
  h_{22}(\chi) &=&2H_0M_{Pl}^2\alpha \lambda_0\left[ \sin\chi+\chi\cos\chi \right]\exp (-\alpha \chi \sin(\chi)),
\end{eqnarray}
where $\psi= t+\psi_*,~~\chi=t+\chi_*$.

The potential $V(\psi,\chi)$ is
\begin{equation}\nonumber
 V(\psi,\chi)= 3H_0^2 M_{Pl}^2\left[ (1-\lambda_0)\exp (-\alpha \psi \sin \psi)+\lambda_0\exp (-\alpha \chi \sin \chi)\right]^2 -
\end{equation}
\begin{equation}\label{V-exp}
{}-H_0 M_{Pl}^2\alpha\left[(1-\lambda_0)(\sin \psi+\psi \cos\psi) \exp (-\alpha \psi \sin \psi)+\lambda_0(\sin \chi+\chi \cos\chi)\exp (-\alpha \chi \sin \chi)\right].
\end{equation}

\subsection{Example of an hyperbolic Hubble function}

Another example is connected with the scale factor
 \begin{equation*}
 a(t)=a_0[k_* \sinh (\lambda t)]^{2/3},
  \end{equation*}
which  corresponds to the $\Lambda$CDM model. The Hubble function is
 \begin{equation*}
 H=H_0\coth (\lambda t),\quad H_0=\frac{2}{3}\lambda.
 \end{equation*}
The corresponding superpotential is
\begin{equation*}
W(\psi,\chi)=H_0\left[ (1-\lambda_0)\coth (\lambda\psi)+\lambda_0\coth (\lambda\chi)\right]
\end{equation*}

 The solution is
 \begin{eqnarray}
   \psi &=& t +\psi_*\\
  \chi &=&  t+\chi_* \\
   h_{11}(\psi) &=& 2M_{Pl}^2H_0(1-\lambda_0)\lambda\sinh^{-2}(\lambda\psi )\\
   h_{22}(\chi) &=&  2M_{Pl}^2H_0\lambda_0\lambda\sinh^{-2}(\lambda\chi)
 \end{eqnarray}
Here and further we assume that $ \psi \rightarrow (\psi-\psi_*)$ and $ \chi \rightarrow (\chi-\chi_*)$.
The physical potential is
\begin{equation}\nonumber
V(\psi,\chi)= 3M_{Pl}^2H_0^2\left[(1-\lambda_0)\coth(\lambda\psi) +\lambda_0\coth(\lambda\chi)\right]^2 -
\end{equation}
\begin{equation}\label{rec-6}
-M_{Pl}^2H_0\lambda \left[(1-\lambda_0)\sinh^{-2}(\lambda\psi) +\lambda_0\sinh^{-2}(\lambda\chi)\right].
\end{equation}

Using the superpotential method, it is possible to get different models for the given time dependence of the Hubble parameter, which we demonstrate in the next section.

\section{The cyclic Universe}\label{QSSCM}
In this section, we focus on a cyclic type of Universe dubbed Quasi-Steady State (QSS) introduced to address the outstanding problems of the hot big bang, for instance, the singularity problem, in particular, see Refs.~\cite{Hoyle,Banerjee,Sachs} and the references therein. In this case, the Hubble parameter might increase at late times to account the well-known tension between  Planck and local observations. The proposed early Universe modifications, namely, the interaction between known matter components or their interaction with dark energy, do not seem to account for the discrepancy~\cite{Reis}. One might attribute the latter to late time physics, for instance,  to the emergence of phantom behaviour at late times. The quasi-steady state model includes such a feature. In what follows, we construct single- (double-) field potentials corresponding to the QSS.

\subsection{One-field models}

Let us construct a CCM with the following form of scale factor that characterizes the quasi-steady state theory,
\begin{equation}\label{qsa}
a(t)=a_0\mathrm{e}^{ C_W  t}(1+\alpha \cos (\mu t)),
\end{equation}
where $a_0$, $\alpha$, and $\mu$ are constants.
This type of  dynamics corresponds to the quasi-steady state model~\cite{Hoyle,Banerjee,Sachs}.
We assume that $a(t)>0$ for any values of $t$, so $|\alpha|<1$. The shift of time $t\rightarrow t+\pi/\mu$ is equivalent to the change of the sign of $\alpha$, so we can assume that $0\leqslant\alpha<1$ without loss of generality\footnote{The case $\alpha=0$ corresponds to a de Sitter solution.}.   For the same reason, we can put $\mu>0$. The corresponding Hubble parameter is
\begin{equation}
\label{H1}
    H= C_W  -\frac{\mu\alpha\sin\left(\mu t\right)}{1+\alpha\cos\left(\mu t\right)},
\end{equation}
and it has the following time derivative:
\begin{equation}
\label{dH1}
   \dot H={}-\mu^2\alpha\frac{\alpha+\cos\left(\mu t\right)}{\left[1+\alpha\cos\left(\mu t\right)\right]^2}.
\end{equation}

Such behavior of the Hubble parameter can be reproduced in one-field models.
Reconstructing the chiral metric component as in the previous section, we choose $\psi=\mu t$ and get
\begin{equation}
h_{11}=\frac{2\alpha (\cos \phi +\alpha)}{(1+\alpha\cos \phi)^2}
\end{equation}
and
\begin{equation}
V(\phi)=\frac{ C_W ^2+\alpha^2( C_W ^2-\mu^2)\cos^2\phi +\alpha\cos\phi (2 C_W ^2-\mu^2)-2\alpha C_W \mu\sin\phi(1-\alpha\cos\phi)}{(1+\alpha\cos\phi)^2 }\,.
\end{equation}

The alternative variant is to consider $\psi(t)$ that tends to finite limits at $t\rightarrow\pm\infty$:
\begin{equation}
\label{psi}
    \psi(t)=\frac{2}{\mu\sqrt{1-\alpha^2}}\arctan\left(\frac{\sqrt{1-\alpha^2}}{1+\alpha}\tan\left(\frac{\mu t}{2}\right)\right).
\end{equation}

It is easy to check that
\begin{equation}
\label{cos}
\cos^2\left(\frac{\mu t}{2}\right)=\frac{(1-\alpha)\cos^2\left(\frac{\mu\sqrt{1-\alpha^2}}{2}\psi \right)}{1+\alpha-2\alpha\cos^2\left(\frac{\mu\sqrt{1-\alpha^2}}{2}\psi \right)},\quad\Rightarrow\quad
\cos\left(\mu t\right)=\frac{\alpha-\cos\left(\mu\sqrt{1-\alpha^2}\psi \right)}{\alpha\cos\left(\mu\sqrt{1-\alpha^2}\psi \right)-1}
\end{equation}

Thus, the function $\psi(t)$ is a solution of the following equation:
\begin{equation}
\label{equpsi}
    \dot\psi=\frac{1}{1+\alpha\cos\left(\mu t\right)}=U(\psi)=\frac{1-\alpha\cos\left(\mu\sqrt{1-\alpha^2}\psi \right)}{1-\alpha^2}\,.
\end{equation}

In the case of a one scalar field model
\begin{equation}
    \dot H={}-\frac{1}{2M_{Pl}^2}h_{11}(\psi){\dot{\psi}}^2,
\end{equation}
therefore,
\begin{equation*}
    h_{11}=2M_{Pl}^2\mu^2\left[\alpha^2+\alpha\cos\left({\mu t}\right)\right]=2M_{Pl}^2\mu^2\alpha\left[   \frac{(1-\alpha^2)\cos\left(\mu\sqrt{1-\alpha^2}\psi \right)}{1-\alpha\cos\left(\mu\sqrt{1-\alpha^2}\psi \right)}\right].
\end{equation*}

Using Eq.~(\ref{WF2}), we obtain
\begin{equation}\label{dWpsi}
    W'_{,\psi}=\mu^2\alpha\cos\left(\mu\sqrt{1-\alpha^2}
    \psi\right) \quad \Rightarrow \quad
    W=\frac{\mu\alpha}{\sqrt{1-\alpha^2}}\sin\left(\mu\sqrt{1-\alpha^2}\psi\right).
\end{equation}
The potential of the model constructed is defined by (\ref{e5-2}) and has the following form:
\begin{equation}
\label{Vpsi1}
\begin{split}
    V &{} = \frac{M_{Pl}^2\mu^2\alpha}{1-\alpha^2}\left[3\alpha-\cos\left(\mu\sqrt{1-\alpha^2}\psi\right)
   -2\alpha\cos^2\left(\mu\sqrt{1-\alpha^2}\psi\right)\right]\\
   &{}-\frac{6\alpha\mu C_W   M_{Pl}^2}{\sqrt{1-\alpha^2}}\sin\left(\mu\sqrt{1-\alpha^2}\psi\right)+
   3M_{Pl}^2 C_W ^2.
 \end{split}
\end{equation}

Note that $h_{11}$ changes its sign during the scalar field evolution, so $\psi$ is neither an ordinary scalar field nor a phantom scalar field~\cite{sami,Nojiri:2003vn}. Assuming that the considering one-field models describe the dark energy, we get the result that the obtained exact solutions have the state parameters crossing the cosmological constant barrier. It has been shown in Ref.~\cite{Vikman} that such transitions are physically implausible in one-field models because they are either realized by a discrete set of trajectories in the phase space or are unstable with respect to the cosmological perturbations. To describe such a type of dark energy one can use quintom models~\cite{QuntomREV,AKV2,Vernov06,Nesseris:2006er,Lazkoz,Setare,ENO}.

\subsection{The CCM quasi-steady state models}

To construct a two-field CCM with the Hubble parameter given by (\ref{H1}) we use relations
$W_{,\psi}=-\frac{1}{2}h_{11}(\psi)\dot{\psi}$, $W_{,\chi}=-\frac{1}{2}h_{22}(\chi)\dot{\chi}$, and
\begin{equation}\label{dotH-1}
\frac{dW}{dt}={}-\frac{1}{2}h_{11}(\psi)\dot{\psi}^2 -\frac{1}{2}h_{22}(\chi)\dot{\chi}^2.
\end{equation}

First of all we can easily make a reconstruction of the chiral metric component as in the previous section.

We choose the superpotential in the form
\begin{equation}\label{w-sim}
W(\psi,\chi)={}-(1-\lambda_0)\frac{\alpha\mu\cos \mu t(\psi) }{(1+\alpha\cos \mu t(\psi))^2}-\lambda_0\frac{\alpha\mu\cos \mu t(\chi) }{(1+\alpha\cos \mu t(\chi))^2}
\end{equation}

Then we choose the dependence on $t$ as
\begin{equation}\label{exm-12}
\psi=t +\psi_*,~~ \chi=t+\chi_*
\end{equation}

The chiral metric components will be
\begin{equation}\label{exm-h1}
h_{11}(\psi)=\frac{2\alpha\mu^2\cos \mu \psi }{(1+\alpha\cos \mu \psi)^2}
\end{equation}
\begin{equation}\label{exm-h2}
h_{22}(\chi)=\frac{2\alpha\mu^2}{(1+\alpha\cos \mu\chi)^2}
\end{equation}

The physical potential can be easily derived by Eq.~\eqref{V-W}.
(It can be written here, but it is rather large.)

The superpotential is evidently defined by \eqref{w-sim} with the substitution \eqref{exm-12}.

There are other possible solutions connecting with the choice of chiral metric components. Let us study a few of them.

From Eq.~\eqref{w-sim}, we find that
\begin{equation}\label{W-dot}
\frac{dW}{dt} ={}-\frac{\alpha\mu^2}{(1+\alpha\cos (\mu t))^2}(\cos (\mu t)+\alpha).
\end{equation}

Further we can decompose the expression above into the following two:
\begin{equation}\label{W-dot-1}
\frac{\partial W(\psi,\chi)}{\partial \psi}={}-\frac{\alpha\mu^2\cos \mu t }{(1+\alpha\cos \mu t)^2}={}-\frac{1}{2}h_{11}(\psi)\dot{\psi}^2
\end{equation}
\begin{equation}\label{W-dot-2}
\frac{\partial W(\psi,\chi)}{\partial \chi}={}-\frac{\alpha^2\mu^2}{(1+\alpha\cos \mu t)^2}={}-\frac{1}{2}h_{22}(\chi)\dot{\chi}^2
\end{equation}

Using  relations \eqref{exm-12}, we can perform integration of the Eqs.~\eqref{W-dot-1} and \eqref{W-dot-2} and obtain two parts of the superpotential $W_1(\psi)$ and $W_2(\chi)$ in terms of elementary functions, but in rather complicated form.
Therefore, our task is to make a combination which allows to integrate the expressions for the fields and for the superpotential with a more suitable result.

To this end, we can choose
\begin{equation}\label{i-1}
\dot{\psi}^2=\frac{\mu^2}{(1+\alpha\cos \mu t)^2},~~h_{11}=2\alpha\cos \mu t
\end{equation}
\begin{equation}\label{i-2}
\dot{\chi}^2=\frac{\mu^2}{(1+\alpha\cos \mu t)^2},~~h_{22}=2\alpha^2
\end{equation}

As the result, we obtain the following solution for the chiral fields:
\begin{equation}
\label{psi-1}
\psi(t)=\frac{2}{\sqrt{1-\alpha^2}}\arctan\left(\frac{1-\alpha}{\sqrt{1-\alpha^2}}
\tan\left(\frac{\mu t}{2}\right)\right).
\end{equation}
\begin{equation}
\label{chi-1}
  \chi(t)=\frac{2}{\sqrt{1-\alpha^2}}\arctan\left(\frac{1-\alpha}{\sqrt{1-\alpha^2}}
\tan\left(\frac{\mu t}{2}\right)\right).
\end{equation}

From solution \eqref{psi-1} one can obtain the chiral metric component $h_{11}(\psi)$ in the form
\begin{equation}\label{h22-psi}
h_{11}(\psi)=2\alpha\frac{(1-\alpha)-(1+\alpha)\tan^2\left(\frac{\sqrt{1-\alpha^2}}{2}\psi\right)}
{(1-\alpha)+(1+\alpha)\tan^2\left(\frac{\sqrt{1-\alpha^2}}{2}\psi\right)}
\end{equation}

Once again such a presentation does not give for us a suitable form for the superpotential and for physical potential.

For the same representation \eqref{i-1} one can choose another appearance for $\chi$ and $h_{22}$. For example,
\begin{equation}\label{i-3}
\dot{\chi}^2=2\alpha^2,~~h_{22}=\frac{\mu^2}{(1+\alpha\cos \mu t)^2}.
\end{equation}
Then the solution for $\chi $ is
\begin{equation*}
\chi=\sqrt{2}\alpha t +\chi_*
\end{equation*}
and for $h_{22}(\chi)$,
\begin{equation*}
h_{22}(\chi)=\frac{\mu^2}{\left(1+\alpha\cos\left( \frac{\mu}{\sqrt{2}\alpha}\chi\right)\right)^2}.
\end{equation*}

Thus, we can state that the chiral metric, the fields, and the physical potential may be different with respect to the given Hubble function.
To stress this fact and to find the suitable form of the superpotential we generalize the procedure of solutions generating by introducing two arbitrary functions $f(\mu t)$ and $y(\mu t)$ in the following way:
\begin{equation}\label{ii-1}
\alpha\mu^2\frac{f(\mu t)^2\cos \mu t }{f(\mu t)^2(1+\alpha\cos \mu t)^2}=\frac{1}{2}h_{11}(\psi)\dot{\psi}^2
\end{equation}
\begin{equation}\label{ii-2}
\alpha^2\mu^2\frac{y(\mu t)^2}{y(\mu t)^2(1+\alpha\cos \mu t)^2}=\frac{1}{2}h_{22}(\chi)\dot{\chi}^2
\end{equation}

The functions $f(\mu t)$ and $y(\mu t)$ can be selected in such a way that integration is performed while finding fields. For example, if we are finding the field $\psi$ we have to perform the integral
\begin{equation}\label{iii-1}
\psi=\int \frac{d(\mu t) }{f(\mu t)(1+\alpha\cos \mu t)}
\end{equation}
To make the integral with an elementary function a solution one can choose as an example $f(\mu t)=\frac{1}{\sin \mu t}$. Then the solution is
\begin{equation}\label{iii-2}
 \psi={}-\frac{1}{\alpha}\ln (1+\alpha\cos \mu t),\quad \mbox{where}\quad \alpha<1.
\end{equation}

Under suggestion \eqref{iii-1}, one can find the chiral metric component $h_{11}$ in the following way
\begin{equation}\label{iii-3}
h_{11}(\psi)=2\alpha f(\mu t)^2\cos \mu t\,.
\end{equation}
In our case, $h_{11}=2\alpha\frac{\cos \mu t}{\sin^2\mu t}$.
Finding dependence $t$ on $\psi$ from \eqref{iii-2}, one can obtain dependence
$h_{11}$ on $\psi$ as follows
\begin{equation}\label{iii-4}
  h_{11}(\psi)=2\alpha^2 \frac{e^{-\alpha\psi}-1}{\alpha^2-(e^{-\alpha\psi}-1)^2}
\end{equation}

Further we consider which types of  superpotential and physical potential will correspond to the solution for field $\psi$ \eqref{iii-2} and chiral metric component  $h_{11}(\psi)$ \eqref{iii-4} using the freedom of the possible choice of function $y(\chi)$.

Let us start with the solution for the $\psi$-part in the form ($|\alpha|<1$)
\begin{equation}\label{iii-1-1}
 \psi=-\frac{1}{\alpha}\ln (1+\alpha\cos \mu t),\quad~~f(\mu t)=\frac{1}{\sin \mu t},
\end{equation}
\begin{equation}\label{iii-4-1}
  h_{11}=2\alpha^2 \frac{e^{-\alpha\psi}-1}{\alpha^2-(e^{-\alpha\psi}-1)^2}\,.
\end{equation}
Note that
\begin{equation*}
\dot{\psi}(\psi)=\frac{\mu}{\alpha}e^{\alpha\psi}\sqrt{\alpha^2-(e^{-\alpha\psi}-1)^2}\,.
\end{equation*}

To find the dependence $W$ on $\psi$ we have to integrate the relation
\begin{equation*}
W_{,\psi}=-\frac{1}{2}h_{11}(\psi)\dot{\psi},
\end{equation*}
which can be transformed to the following:
\begin{equation}\label{w-h11}
\frac{\partial W}{\partial \psi}={}-\mu\alpha e^{-\alpha\psi}\frac{e^{-\alpha\psi}-1}{\sqrt{\alpha^2-(e^{- \alpha\psi}-1)^2}}\,.
\end{equation}

The solution is
\begin{equation}\label{w-psi-1}
 W_1(\psi)={}-\frac{\mu}{a^2}F(z)+\frac{\mu\alpha^2}{a^3}\arctan \left( \frac{a^2z-1}{aF(z)}\right)\,.
\end{equation}
where
$$
z=\exp (\alpha\psi),~~F(z)=\sqrt{-a^2z^2+2z-1},~~ a^2=1-\alpha^2>0.
$$

 Further we have various possibilities to define $\chi$ and $h_{22}(\chi)$. Let us select the nontrivial choice $y(\mu t)=1-\alpha\cos \mu t$. Then we have the solution for the field  $\chi$,
 $$
 \chi=\frac{1}{\sqrt{1-\alpha^2}}\arctan \left[ \frac{\tan \mu t}{\sqrt{1-\alpha^2}}\right],~~\alpha^2<1.
 $$

The chiral metric component $h_{22}$ is
 \begin{equation*}
h_{22}(\chi)=2\alpha^2\left( 1-\alpha \left(1+\left(1-\alpha^2\right)\tan^2\left(\sqrt{1-\alpha^2}\chi\right)\right)^{-1/2}\right)^2.
\end{equation*}

Thus,  we  get
\begin{equation}\label{DWchi}
\frac{dW(\chi)}{d\chi}={}-\frac{\alpha^2\mu}{(1-\alpha^2)^{3/2}} \left(\sqrt{1+(1-\alpha^2)v^2}-\alpha\right)^2\left(1+v^2\right)^{-2},\quad~v=\tan\left(\sqrt{1-\alpha^2}\chi\right).
\end{equation}

 The solution for the $\chi$ part of the superpotential is
\begin{equation}\nonumber
 W_2(\chi)={}-\frac{\alpha\mu}{(1-\alpha^2)^{2/3}}\left\{ 2(\alpha^2-1)\arctan\left( v/P(v)\right) -(2\alpha^2-1)\arctan\left( \alpha v/P(v)\right)+\right.
 \end{equation}
 \begin{equation}\label{Wchi}
\left. \frac{\alpha v(\alpha+1)P(v)}{1+v^2}-\frac{\sqrt{1-\alpha^2}}{2}\chi-\frac{\alpha^2v}{v^2+1} \right\},~~P(v)= \sqrt{v^2(1-\alpha^2)+1}\,.
 \end{equation}

The superpotential is equal to
\begin{equation*}
W(\psi,\chi)= W_1(\psi)+W_2(\chi).
\end{equation*}

The next step is to calculate the potential $V(\psi,\chi)$ by Eq.~(\ref{V-W}). This is possible but the answer will be too long.

It is possible to get the same time evolution of the scalar factor using a quintom model.
The time derivative of the Hubble parameter (\ref{dH1}) can be presented in the following form:
\begin{equation}
\label{dHH}
   \dot H={}-\frac{\mu^2\alpha\left(\alpha+\cos\left(\mu t\right)\right)}{\left[1+\alpha\cos\left(\mu t\right)\right]^2}=\frac{-(\alpha^2+\alpha)\mu^2}
    {\left[1+\alpha\cos\left(\mu t\right)\right]^2}+
    \frac{2\mu^2\alpha\sin^2\left(\frac{\mu t}{2}\right)}{\left[1-\alpha+2\alpha\cos^2\left(\frac{\mu t}{2}\right)\right]^2}.
\end{equation}

Let us introduce two scalar fields with the same time behavior:
\begin{equation}
\label{psichi}
\begin{split}
\psi(t)&=\frac{2}{\mu\sqrt{1-\alpha^2}}\arctan\left(\frac{\sqrt{1-\alpha^2}}{1+\alpha}\tan\left(\frac{\mu t}{2}\right)\right),\\
 \chi(t)&=\frac{2}{\mu\sqrt{1-\alpha^2}}\arctan\left(\frac{\sqrt{1-\alpha^2}}{1+\alpha}\tan\left(\frac{\mu t}{2}\right)\right).
\end{split}
\end{equation}

Using Eq.~(\ref{cos}), the time derivative of $H$ can be rewritten as follows:
\begin{equation}\label{dHHf}
 \dot H= {}-(1+\alpha)\alpha\mu^2{\dot{\psi}}^2+2\alpha\mu^2\left[1-\frac{(1-\alpha)\cos^2\left(\frac{\mu\sqrt{1-\alpha^2}}{2}\chi \right)}{1+\alpha-2\alpha\cos^2\left(\frac{\mu\sqrt{1-\alpha^2}}{2}\chi \right)}\right]{\dot{\chi}}^2\,.
\end{equation}
So, we get the the matrix $h_{AB}$ as diagonal and
\begin{equation}\label{hmatrix}
    h_{11}=2(1+\alpha)\alpha M_{Pl}^2\mu^2,\qquad h_{22}={}-\frac{2\alpha(1+\alpha)\mu^2 M_{Pl}^2\left[1-\cos\left(\mu\sqrt{1-\alpha^2}\chi \right)\right]}{1-\alpha\cos\left(\mu\sqrt{1-\alpha^2}\chi \right)}\,.
\end{equation}
The field $\psi(t)$ is an ordinary scalar field, whereas $\chi(t)$ is a phantom scalar field.

We assume that the superpotential has the form~(\ref{Wsum}):
\begin{equation}\label{WH}
    W= W_1(\psi)+W_2(\chi).
\end{equation}

Using Eq.~(\ref{WF2}), we get
\begin{equation}
\label{W1psi}
    W_1=\frac{\mu^2\alpha}{\alpha-1}\psi-\frac{\mu\alpha^2}{\sqrt{1-\alpha^2}(\alpha-1)} \sin\left(\mu\sqrt{1-\alpha^2}\psi\right),
\end{equation}
\begin{equation}
\label{W1chi}
    W_2=\frac{\mu^2\alpha}{\alpha-1}\chi-\frac{\mu\alpha}{\sqrt{1-\alpha^2}(\alpha-1)}\sin\left(\mu\sqrt{1-\alpha^2}\chi\right).
\end{equation}

Using Eq.~(\ref{V-W}), we get the potential of the obtained two-field model in the following form:
\begin{equation}
\label{Vpsi}
\begin{split}
    V &{} = \frac{3M_{Pl}^2}{\left(\alpha-1\right)^3\left(\alpha+1\right)}
    \left[\alpha^2\mu^4(\alpha^2-1)\left(\psi^2+\chi^2\right)+ 2\mu^4\alpha^2(\alpha^2-1)\psi\chi\right.\\
    &{}+2\mu^2\alpha C_W (\alpha+1)(\alpha-1)^2\psi+
    2\mu^2\alpha C_W (\alpha+1)(\alpha-1)^2\chi+(\alpha+1)(\alpha-1)^3 C_W ^2\\
     &{}-\alpha^2(\alpha^2+1)\mu^2-2\alpha^3\mu^2\sin\left(\mu\sqrt{1-\alpha^2}\psi\right)\sin\left(\mu\sqrt{1-\alpha^2}\chi\right)\\
     &{}+ 2\alpha\mu\sqrt{1-\alpha^2}\left[\alpha\mu^2(\chi+\psi)+(\alpha-1) C_W \right]\sin\left(\mu\sqrt{1-\alpha^2}\chi\right)\\
     &{}+2\alpha^2\mu\sqrt{1-\alpha^2}\left[\alpha\mu^2(\chi+\psi)+(\alpha-1) C_W \right]\sin\left(\mu\sqrt{1-\alpha^2}\psi\right)\\
     &{}+\frac{1}{3}\mu^2\alpha^2\left[(\alpha+2)\cos\left(\mu\sqrt{1-\alpha^2}\chi\right)^2+(2\alpha^2+\alpha)
     \cos\left(\mu\sqrt{1-\alpha^2}\psi\right)^2\right]\\
          &\left.+{}\frac{1}{3}\mu^2\left(\alpha(1-\alpha^2)\cos\left(\mu\sqrt{1-\alpha^2}\chi\right)
          +2\alpha^2(\alpha-1)\cos\left(\mu\sqrt{1-\alpha^2}\psi\right)\right)  \right].
 \end{split}
\end{equation}

\section{Reducing two-field dynamic equations to the single-field ones}
\label{2to1}
For simplification of the generation of the exact solutions for CCMs we consider the possibility
of reducing the dynamic equations in such a type of models to a single-field case.

For this aim, we write the dynamic equations in terms of the effective field $\varphi$, which is connected with CCM fields ${\phi}^{A}$
by the following relation \cite{chervon26}:
\begin{equation}
\label{effectiveF}
\dot{\varphi}^{2}=h_{AB}\dot{\phi}^{A}\dot{\phi}^{B},
\end{equation}
where $\phi^{A}$ are the fields of the CCM, and $h_{AB}$ is the metric tensor of a target space.

The double kinetic energy $X=\dot{\varphi}^{2}$ of the effective field $\varphi$ was considered earlier as $X$ field (\ref{chi}).
For positive $X=\dot{\varphi}^{2}>0$ one has the canonical effective scalar field $\varphi$ and $X<0$ corresponds to the phantom one.  The dynamic equations in CCMs (\ref{a2})--(\ref{f-5m}) can be noted as
\begin{equation}
\label{CCM1}
3H^{2}M^{2}_{Pl}=\frac{1}{2}\dot{\varphi}^{2}+V(\varphi),
\end{equation}
\begin{equation}
\label{CCM2}
\dot{\varphi}^{2}=-2\dot{H}M^{2}_{Pl},
\end{equation}
\begin{equation}
\label{CCM3}
D_{t}\dot{\phi}^{A}+3H\dot{\phi}^{A}+h^{AB}V_{,B}=0,
\end{equation}
where
\begin{equation}
D_{t}\dot{\phi}^{A}=\frac{d\dot{\phi}^{A}}{dt}+\Gamma^{A}_{BC}\dot{\phi}^{B}\dot{\phi}^{C}
\end{equation}
is a covariant derivative in a target space.

Also, one can rewrite the first dynamic equation (\ref{CCM1}) on the basis of Eq.~(\ref{CCM2}) in the following form:
\begin{equation}
V(\varphi)=M^{2}_{Pl}(3H^{2}+\dot{H}).
\end{equation}

Therefore, in the general case, the connection between particular solutions of CCMs and one-field models is defined on the basis of Eq.~(\ref{CCM3}) as follows:
\begin{eqnarray}
\label{connection}
\ddot{\phi}^{A}+\Gamma^{A}_{BC}\dot{\phi}^{B}\dot{\phi}^{C}+3H\dot{\phi}^{A}+h^{AB}V_{,B}=0, \quad \Leftrightarrow \quad
\ddot{\varphi}+3H\dot{\varphi}+V_{,\varphi}=0,
\end{eqnarray}
for all $A$.

Further we consider the fulfillment of this condition for the case of a CCM with two identical scalar fields $\chi=\psi$ by the specific connections between components of the tensor $h_{AB}$.

\subsection{The CCM with two scalar fields}

Now, we consider the partial case of a CCM with two identical scalar fields $\chi=\psi$ and the specific connections between components of the tensor $h_{AB}$.

Firstly, we write the dynamic equations (\ref{CCM1})--(\ref{CCM3}) for a CCM with two fields:
\begin{equation}
\label{EC1}
3H^2M^{2}_{Pl}=\frac{1}{2}h_{11}\dot\chi^2 +h_{12}\dot\chi \dot\psi+
\frac{1}{2}h_{22}\dot\psi^2 +V(\psi,\chi),
\end{equation}
\begin{equation}
\label{EC2}
-\dot{H}M^{2}_{Pl} =\frac{1}{2}h_{11}\dot\chi^2
+h_{12}\dot\chi \dot\psi+\frac{1}{2}h_{22}\dot\psi^2,
\end{equation}
\begin{eqnarray}
\label{F1}
\nonumber
3H\left( h_{11}\dot\chi +h_{12}\dot\psi \right) + \frac{\partial }{\partial t}\left( h_{11}\dot\chi
+h_{12}\dot\psi \right)-\frac{1}{2}\frac{\partial h_{11}}{\partial \chi}\dot\chi^2 -\\
\frac{\partial h_{12}}{\partial \chi}\dot\chi \dot\psi
-\frac{1}{2}\frac{\partial h_{22}}{\partial \chi}\dot\psi^2 +\frac{\partial V}{\partial\chi}=0,
\end{eqnarray}
\begin{eqnarray}
\label{F2}
\nonumber
3H\left( h_{12}\dot\chi +h_{22}\dot\psi \right)  + \frac{\partial }{\partial t}\left( h_{12}\dot\chi
+h_{22}\dot\psi \right)
 -\frac{1}{2}\frac{\partial h_{11}}{\partial \psi}\dot\chi^2 -\\
\frac{\partial h_{12}}{\partial \psi}\dot\chi \dot\psi
-\frac{1}{2}\frac{\partial h_{22}}{\partial \psi}\dot\psi^2 +\frac{\partial V}{\partial\psi}=0.
\end{eqnarray}

Secondly, for models with the following metric tensor of the target space,
\begin{equation}\label{hAB}
h_{AB}=\begin{pmatrix}
\frac{n}{2}-h_{12}(\psi,\chi)& h_{12}(\psi,\chi)\\
h_{12}(\psi,\chi) & \frac{n}{2}-h_{12}(\psi,\chi)\\
\end{pmatrix}
\end{equation}
where $h_{11}=h_{22}=\frac{n}{2}-h_{12}$, $h_{21}=h_{12}$ and $n$ is an arbitrary constant, under condition $\chi=\psi$, we have
\begin{equation}
 h_{11}\dot\chi +h_{12}\dot\psi=h_{12}\dot\chi +h_{22}\dot\psi=\frac{n}{2}\dot\chi=\frac{n}{2}\dot\psi,
\end{equation}
\begin{equation}
\frac{1}{2}\frac{\partial h_{11}}{\partial \chi}\dot\chi^2+\frac{\partial h_{12}}{\partial \chi}\dot\chi \dot\psi
+\frac{1}{2}\frac{\partial h_{22}}{\partial \chi}\dot\psi^2=\frac{1}{2}\frac{\partial h_{11}}{\partial \psi}\dot\chi^2+\frac{\partial h_{12}}{\partial \psi}\dot\chi \dot\psi +\frac{1}{2}\frac{\partial h_{22}}{\partial \psi}\dot\psi^2=0,
\end{equation}
\begin{equation}
\frac{1}{2}h_{11}\dot\chi^2 +h_{12}\dot\chi \dot\psi+\frac{1}{2}h_{22}\dot\psi^2=
\frac{n}{2}\dot\chi^{2}=\frac{n}{2}\dot\psi^{2}.
\end{equation}

Thus, from Eqs.~(\ref{EC1})--(\ref{F2}) we obtain
\begin{equation}
\label{E1}
V(\psi,\chi)=M^{2}_{Pl}(3H^2+\dot{H})=V(\varphi),
\end{equation}
\begin{equation}
\label{E2}
-\dot{H}M^{2}_{Pl} =\frac{n}{2}\dot\chi^2=\frac{n}{2}\dot\psi^2=\frac{1}{2}\dot{\varphi}^{2},
\end{equation}
\begin{equation}
\label{E3}
\ddot{\chi}+3H\dot\chi +\frac{2}{n}\frac{\partial V}{\partial\chi}=\ddot{\psi}+3H\dot\psi+\frac{2}{n}\frac{\partial V}{\partial\psi}=0.
\end{equation}

From Eq. (\ref{E3}), taking into account the equality of the scalar fields $\chi=\psi$, we have the condition of symmetry of the potential $\frac{\partial V}{\partial\chi}=\frac{\partial V}{\partial\psi}$ with respect to these fields.

Therefore, one can write
\begin{equation}
\label{SymCon}
\frac{dV}{d\varphi}=\frac{\partial V}{\partial\chi}\frac{d\chi}{d\varphi}+\frac{\partial V}{\partial\psi}\frac{d\psi}{d\varphi}=
2\frac{\partial V}{\partial\chi}\frac{d\chi}{d\varphi}=2\frac{\partial V}{\partial\psi}\frac{d\psi}{d\varphi}\,.
\end{equation}

Finally, we note that from (\ref{E1})--(\ref{E2}) one can obtain  (\ref{E3}) and the equation
\begin{equation}
\label{eqfield4}
\ddot{\varphi}+3H\dot{\varphi}+V_{,\varphi}=0,
\end{equation}
is a differential consequence of (\ref{E1})--(\ref{E2}) as well, which can easily be  obtained by substituting the effective field
\begin{equation}
\label{E3CF}
\varphi=\pm\frac{\sqrt{n}}{2}\left(\chi+\psi\right),
\end{equation}
in the form $\varphi=\pm\sqrt{n}\chi$ or $\varphi=\pm\sqrt{n}\psi$ into Eq.~(\ref{eqfield4}) with using the relations (\ref{SymCon}).

Therefore, in this case, we have the fulfillment of the condition (\ref{connection}), that allows one to reduce the initial CCM with two scalar fields to the one-field model.

Further we consider the following superpotential of the effective field
\begin{equation}
\label{SUPHT}
W(\varphi)\equiv H(t),
\end{equation}
and write the dynamic equations (\ref{E1})--(\ref{E3}) as
\begin{equation}
\label{SOL1}
V(\varphi)=M^{2}_{Pl}\left[3W^{2}(\varphi)-2M^{2}_{Pl}\left(\frac{dW(\varphi)}{d\varphi}\right)^{2}\right],
\end{equation}
\begin{equation}
\label{SOL2}
\dot{\varphi}={}-2M^{2}_{Pl}\left(\frac{dW(\varphi)}{d\varphi}\right),
\end{equation}
for the effective field (\ref{E3CF}), where the constant parameter $n$ defines the character of an effective field $\varphi$, namely, this field can be canonical or phantom for the different signs of $n$.

Thus, we have a connection between chiral cosmological models with two canonical scalar fields $\chi$ and $\psi$ and a single-field model with a effective field $\varphi$. The sign on the parameter $n$ depends on the choice of field $\varphi$, for a canonical effective field one has $n>0$, for a phantom one the parameter $n<0$.

As an example of the proposed approach, we will obtain the exact solutions for two identical scalar fields $\chi$ and $\psi$ with linear dependence from cosmic time for an arbitrary function $h_{12}(\chi,\psi)$ in the metric tensor of the target space (\ref{hAB}) by choosing the special form of the superpotential $W(\varphi)$. Thus, these solutions will differ from ones considered earlier in Section~VII and Section~VIII  with a specific expressions of the components of the tensor $h_{AB}$ for a given type of the evolution of scalar fields.

For the following superpotential
\begin{equation}
\label{SOLW}
W(\varphi)={}-\frac{\alpha}{2 M^{2}_{Pl}}\varphi,
\end{equation}
from Eq.~(\ref{SOL2}) we have
 \begin{equation}
\varphi(t)=\alpha t-\beta,
\end{equation}
where $\beta$ is the constant of integration.

The Hubble parameter and the scale factor are
\begin{equation}
\label{DEFF}
H(t)={}-\frac{\alpha}{2M^{2}_{Pl}}(\alpha t-\beta),
\end{equation}
\begin{equation}
\label{DEFF1}
a(t)=a_{0}\exp\left[\frac{\alpha}{4M^{2}_{Pl}}t\left(2\beta-\alpha t\right)\right],
\end{equation}
corresponding to Ruzmaikin's solutions~\cite{Ruzmaikina,Skugoreva:2014gka}.

From Eq.~(\ref{SOL1}) we obtain the following potential of the effective field:
\begin{equation}
V(\varphi)=\left(\frac{\alpha}{2M_{Pl}}\right)^{2}\left[3\varphi^{2}-2M^{2}_{Pl}\right].
\end{equation}

After substituting (\ref{E3CF}) into the solutions for $\varphi$, we obtain
\begin{equation}
\label{CCMF1}
\chi(t)=\psi(t)=\pm\frac{1}{\sqrt{n}}\left(\alpha t-\beta\right),
\end{equation}
\begin{equation}
\label{CCMPOT1}
V(\psi,\chi)=\left(\frac{\alpha}{2M_{Pl}}\right)^{2}\left[\frac{3n}{4}(\chi+\psi)^{2}-2M^{2}_{Pl}\right],
\end{equation}
the potential and evolution of the CCM fields corresponding to the same dynamics (\ref{DEFF})--(\ref{DEFF1}) of the early Universe.

For  the other example, we consider the exact solutions defined by the following superpotential
\begin{equation}
\label{HIGGS_SP}
W(\varphi)=\frac{A}{8M^{2}_{Pl}}\varphi^{2}+\lambda,
\end{equation}
where $A$ and $\lambda$ are arbitrary constants.

From Eqs.~(\ref{SOL1}) and (\ref{SOL2}) one has
\begin{equation}
\label{sol1Higgs}
H(t)=B\exp(-At)+\lambda,
\end{equation}
\begin{equation}
\label{sol2Higgs}
a(t)=a_{0}\exp\left(\lambda t-\frac{B}{A}e^{-At}\right),
\end{equation}
\begin{equation}
\label{sol3Higgs}
\varphi(t)=\sqrt{\frac{8B}{A}}\exp\left(-\frac{A}{2}t\right),
\end{equation}
\begin{equation}
\label{sol4Higgs}
V(\varphi)=3\left(\frac{A}{8M_{Pl}}\right)^{2}\varphi^{4}+
\frac{A}{4}\left(3\lambda-\frac{A}{2}\right)\varphi^{2}+3\lambda^{2},
\end{equation}
which correspond to the Higgs potential.

As a special case for $\lambda=A/6$ one has the potential for chaotic inflation
\begin{equation}
V(\varphi)=3\left(\frac{A}{8M_{Pl}}\right)^{2}\varphi^{4}+3\lambda^{2}.
\end{equation}

After replacing the effective field $\varphi$ on CCM fields,
\begin{equation}
\chi(t)=\psi(t)=\pm\sqrt{\frac{8B}{An}}\exp\left(-\frac{A}{2}t\right),
\end{equation}
we have
\begin{equation}
V(\psi,\chi)=\frac{3}{16}\left(\frac{An}{8M_{Pl}}\right)^{2}(\chi+\psi)^{4}+
\frac{nA}{16}\left(3\lambda-\frac{A}{2}\right)(\chi+\psi)^{2}+3\lambda^{2},
\end{equation}
or, for $\lambda=A/6$ the potential of the CCM fields is
\begin{equation}
V(\psi,\chi)=\frac{3}{16}\left(\frac{An}{8M_{Pl}}\right)^{2}(\chi+\psi)^{4}+3\lambda^{2}.
\end{equation}

Under condition $h_{12}(\chi,\psi)=0$ one has the same solutions for the trivial case of a constant diagonal tensor~$h_{AB}$.

Similarly, one can generalize any exact solutions in single-field models (see, for example, \cite{Fomin:2017xlx,Chervon:2017kgn}) on this special class of chiral cosmological models with two components.

\subsection{The generalization of the exact solutions for a CCM with an arbitrary number of fields}

Now, we generalize the proposed method on the case of a CCM with an arbitrary number of interacting similar scalar fields $\phi^{1}(t)=\phi^{2}(t)=...=\phi^{K}(t)$.
In this case, we determine the connection between the diagonal and non-diagonal components of the metric tensor of the target space $h_{AB}$  as
\begin{equation}
\label{COND1}
\sum^{K}_{B=1}h_{CB}=\frac{n}{K},
\end{equation}
for all $C$, with the following condition for non-diagonal components $h_{CB}=h_{BC}$.

Hence, the first  diagonal component is determined as
\begin{equation}
h_{11}=\frac{n}{K}-h_{12}-h_{13}-...-h_{1K},
\end{equation}
and the other components are defined similarly.

In this case, Eqs.~(\ref{CCM1})--(\ref{CCM3}) are reduced to
\begin{equation}
\label{E1M}
V\left(\bar{\phi}\right)=M^{2}_{Pl}(3H^2+\dot{H})=V(\varphi),
\end{equation}
\begin{equation}
\label{E2M}
-\dot{H}M^{2}_{Pl} =\frac{n}{2}\dot\phi^{A}\dot\phi^{A}=\frac{1}{2}\dot{\varphi}^{2},
\end{equation}
\begin{equation}
\label{E3M}
\ddot{\phi}^{A}+3H\dot\phi^{A} +\frac{2}{n}\frac{\partial V(\phi^{A})}{\partial\phi^{A}}=0.
\end{equation}

Therefore, we can generalize the exact solutions (\ref{SUPHT})--(\ref{SOL2}) for the effective field
\begin{equation}
\label{EFFECTIVEFIELD}
\varphi=\pm\frac{\sqrt{n}}{2}\left(\sum^{K}_{A=1}\phi^{A}\right).
\end{equation}

For a CCM with $K$ components, any (similar) field $\phi^A$ can be obtained from the effective field as follows
 \begin{equation}
\label{CCMFIELD}
\phi^A(t)=\pm\frac{2}{K\sqrt{n}}\varphi(t).
\end{equation}

For example, we can generalize the solutions (\ref{SOLW})--(\ref{DEFF1}) for the case of $K$ number of fields:
 \begin{equation}
\phi^{1}(t)=\phi^{2}(t)=...=\phi^{K}(t)=\pm\frac{2}{K\sqrt{n}}\left(\alpha t-\beta\right).
\end{equation}
\begin{equation}
V\left(\bar{\phi}\right)=\left(\frac{\alpha}{2M_{Pl}}\right)^{2}\left[\frac{3n}{4}
\left(\sum^{K}_{A=1}\phi^{A}\right)^{2}-2M^{2}_{Pl}\right].
\end{equation}
In the same way, it is possible to generalize exact solutions for any other models with one scalar field.

Thus, this method gives an integrable class of exact solutions of Eqs.~(\ref{a2})--(\ref{f-1}) for a special case of identical scalar fields and relation (\ref{COND1}) between the components of the metric of the target space.

\section{CONCLUSIONS}

\label{sum}

In this paper, we develop the superpotential technique for chiral cosmological models. The key point in this method is that the Hubble parameter is considered a function of the scalar fields, and this allows one to reconstruct the scalar field potential. The CCM models are actively used in cosmology and can be connected with modified gravity models due to the conformal transformation of the metric. So, the proposed method allows one to construct modified gravity models with exact solutions. In particular, CCMs that correspond to $f(R)$ gravity models with one scalar field have been considered in Section~\ref{SecFR}. Corresponding two-field CCMs with asymptotic de Sitter solutions have been constructed. In the future, we shall explore the possibility of applying  our results to  inflation.

The superpotential method is an effective procedure to construct models with exact particular solutions. In the case of a model with $K$ scalar fields, the superpotential method gives a possibility to get $K$-parametric set of solutions, as we shown in Section~\ref{sp}. In Sections~\ref{SecFR}--\ref{HCpl1t} we found two-parametric sets of exact solutions for two-field CCMs.

To demonstrate that the proposed reconstruction procedure is powerful, we constructed the CCM with different behaviours of the Hubble parameter that are actively used in cosmology. In particular, in Sections~\ref{DIAGCTS} and \ref{HCpl1t} we have found models with Ruzmaikin's solutions that correspond to the intermediate inflation and modified power-law solutions, for which the Hubble parameter is the sum of a constant and a function inverse proportional to the cosmic time. These solutions have been found for models with the given kinetic terms of the actions. In our case, it is possible to choose both the potential and the function that defines the kinetic term. The construction of trigonometric and hyperbolic Hubble functions due to a suitable choice of  kinetic term has been proposed in Section~\ref{TSRec}.

In Section~\ref{QSSCM}, we constructed one- and two-field models that correspond to the Hubble parameter that describes a cyclic type of Universe dubbed quasi-steady state. We demonstrated that the  superpotential method allows one to construct different models with one and the same Hubble parameter.

In Section~\ref{2to1}, we have shown that exact solutions of the CCMs can be obtained by the single-field superpotential method. Comparing this method with the multifield superpotential method developed in Section~\ref{sp}, one can see that the use of the single-field superpotential method allows one to obtain only one-parametric set of solutions, whereas the multifield superpotential method gives rise to  a \emph{K}-parametric set of exact solutions if system~(\ref{FG}) is integrable. Also, the multifield superpotential method is preferable to get exact soluble models with a nonmonotonic Hubble parameter, which corresponds to models with both ordinary and phantom scalar fields. At the same time, for a nonintegrable system~(\ref{FG}), the method proposed in Section~\ref{2to1} is a more simple way to obtain exact solutions.

The  correspondence between one- and multifield models is actively used for multifield inflationary models in the method of cosmological attractors~\cite{KLMattr,MSSM}. Let us note that, compared to the method of cosmological attractors, the proposed algorithm allows one to obtain the exact solutions.
The superpotential method is suitable for construction of inflationary scenarios in one-field models~\cite{Lidsey:1995np,Chervon:2008zz,Binetruy:2014zya,Pieroni:2015cma,Binetruy:2016hna}.
In the future, we plan to generalize this method to the chiral cosmological inflationary models with many scalar fields.

\section*{ACKNOWLEDGEMENTS}

 This work is partially supported by Indo-Russia Project; S.V.C., I.V.F., E.O.P.,  and S.Yu.V. are supported by RFBR Grant No.~18-52-45016  and M.S. is supported by Grant No.~INT/RUS/RFBR/P-315 of the Department of science and technology of India government. S.V.C. is grateful for support by the Program of Competitive Growth of Kazan Federal University.


\begin{thebibliography}{99}

\bibitem{cosmo-obser}
 A.G.~Riess \textit{et al.}  [Supernova Search Team collaboration],
\textit{Type Ia Supernova Discoveries at $z>1$ From the Hubble Space
Telescope: Evidence for Past Deceleration and Constraints on Dark
Energy Evolution},  Astrophys. J. \textbf{607} (2004) 665--687 [arXiv:astro-ph/0402512];\\
 E.~Komatsu, \textit{et al.} [WMAP collaboration],
\textit{Five-Year Wilkinson Microwave Anisotropy Probe (WMAP)
Observations: Cosmological Interpretation},  Astrophys. J.
Suppl. Ser. \textbf{180} (2009) {330--376} [arXiv:0803.0547];\\
W.M. Wood-Vasey {\it et al.} [ESSENCE Collaboration],
\textit{Observational Constraints on the Nature of the Dark Energy:
First Cosmological Results from the ESSENCE Supernova Survey},
Astrophys. J. \textbf{666} (2007) 694--715
[arXiv:astro-ph/0701041]

\bibitem{Bernui:2005pz}
  A.~Bernui, B.~Mota, M.J.~Reboucas, and R.~Tavakol,
  \textit{Mapping large-scale anisotropy in the WMAP data},
  Astron. Astrophys.  {\bf 464} (2007) 479--485 [arXiv:astro-ph/0511666]

\bibitem{Planck2018}
  N.~Aghanim {\it et al.} [Planck Collaboration],
  \textit{Planck 2018 results. VI. Cosmological parameters},
  arXiv:1807.06209 [astro-ph.CO];\\
  Y.~Akrami {\it et al.} [Planck Collaboration],
  \textit{Planck 2018 results. X. Constraints on inflation},
  arXiv:1807.06211 [astro-ph.CO]

\bibitem{Linde}
A.D.~Linde,
   \textit{A New Inflationary Universe Scenario: A Possible Solution of the Horizon, Flatness, Homogeneity, Isotropy and Primordial Monopole Problems},
    Phys.\ Lett. B {\bf 108} (1982)  389;\\
A.D.~Linde,
  \textit{Chaotic Inflation},
  Phys. Lett. B {\bf 129} (1983) 177

\bibitem{nonmin}
B.L. Spokoiny,
\textit{Inflation And Generation Of Perturbations In Broken Symmetric Theory Of Gravity},
Phys. Lett. B \textbf{147}  (1984) 39--43;\\
T. Futamase and K.-i.  Maeda,
\textit{Chaotic Inflationary Scenario In Models Having Nonminimal Coupling With Curvature},
Phys. Rev. D \textbf{39} (1989) 399--404;\\
R. Fakir and W.G. Unruh, \textit{Improvement on cosmological chaotic inflation through nonminimal coupling},
Phys. Rev. D \textbf{41} (1990) 1783--1791;\\
A.O.~Barvinsky and A.Yu.~Kamenshchik, \textit{Quantum scale of inflation and particle physics of the early universe},
Phys. Lett. B {\bf 332} (1994) 270--276 [arXiv:gr-qc/9404062]; \\
J.L.~Cervantes-Cota and H.~Dehnen,
  \textit{Induced gravity inflation in the SU(5) GUT},
  Phys.\ Rev. D {\bf 51} (1995) 395 [arXiv:astro-ph/9412032]; \\
S.~Mukaigawa, T.~Muta and S.D.~Odintsov,
  \emph{Finite grand unified theories and inflation,}
  Int.\ J.\ Mod.\ Phys.\ A {\bf 13} (1998) 2739
  [arXiv:hep-ph/9709299].

\bibitem{SBB1989}
D.S.~Salopek, J.R.~Bond and J.M.~Bardeen,
\textit{Designing Density Fluctuation Spectra in Inflation},
 Phys. Rev. D \textbf{40} (1989) 1753--1788

  \bibitem{SalopekBond}
D.S.~Salopek and J.R.~Bond,
\emph{Nonlinear evolution of long-wavelength metric fluctuations in inflationary models},
Phys. Rev. D \textbf{42} (1990) 3936--3962


\bibitem{Lidsey:1995np}
  J.E.~Lidsey, A.R.~Liddle, E.W.~Kolb, E.J.~Copeland, T.~Barreiro and M.~Abney,
  \textit{Reconstructing the inflation potential: An overview},
  Rev. Mod. Phys. \textbf{69} (1997) 373--410  [arXiv:astro-ph/9508078]

\bibitem{inflation2}
C.M.~Peterson, M.~Tegmark,
\textit{Testing Two-Field Inflation},
Phys. Rev. D \textbf{83} (2011) 023522 [arXiv:1005.4056];\\
Shi Pi, M. Sasaki,
\textit{Curvature perturbation spectrum in two-field inflation with a turning trajectory},
J. Cosmol. Astropart. Phys. \textbf{1210} (2012) 051 [arXiv:1205.0161]

\bibitem{Starobinsky}
 A.A.~Starobinsky,
   \textit{A New Type of Isotropic Cosmological Models Without Singularity},
    Phys.\ Lett. B {\bf 91} (1980)  99--102;\\
A.A.~Starobinsky,
		\emph{Dynamics of phase transition in the new inflationary universe scenario and generation of perturbations,}
		Phys. Lett. B {\bf 117} (1982) 175.

\bibitem{Higgs}
F.L.~Bezrukov and M.~Shaposhnikov,
\textit{The Standard Model Higgs boson as the inflaton},
 Phys. Lett. B \textbf{659} (2008) 703 [arXiv:0710.3755]; \\
A.O.~Barvinsky, A.Y.~Kamenshchik, and A.A.~Starobinsky
\textit{Inflation scenario via the Standard Model Higgs boson and LHC}
{\it J. Cosmol. Astropart. Phys.} {\bf 0811} (2008) 021 [arXiv:0809.2104];\\
 F.~Bezrukov, D.~Gorbunov and M.~Shaposhnikov,
   \textit{On initial conditions for the Hot Big Bang},  J. Cosmol. Astropart. Phys. {\bf 0906} (2009) 029 [arXiv:0812.3622];\\
F.L.~Bezrukov, A.~Magnin and M.~Shaposhnikov,
\textit{Standard Model Higgs boson mass from inflation}
 Phys. Lett. B {\bf 675} (2009) 88 [arXiv:0812.4950];\\
 A.O.~Barvinsky, A.Y.~Kamenshchik, C.~Kiefer, A.A.~Starobinsky,
and C.F.~Steinwachs,
 \textit{Asymptotic freedom in inflationary cosmology with a nonminimally coupled Higgs field},
  	J. Cosmol. Astropart. Phys. \textbf{0912} (2009) 003 [arXiv:0904.1698];\\
F.~Bezrukov   \textit{The Higgs field as an inflaton}  Class. Quant. Grav. {\bf 30} (2013)  214001 [arXiv:1307.0708];\\
A.O.~Barvinsky, A.Y.~Kamenshchik, C.~Kiefer, and C.F.~Steinwachs,
  \textit{The Higgs field as an inflaton Tunneling cosmological state revisited: Origin of inflation with a nonminimally coupled Standard Model Higgs inflaton},
 Phys. Rev. D {\bf 81} (2010) 043530 [arXiv:0911.1408]


\bibitem{Kaiser:2010ps}
  D.I.~Kaiser,
  \emph{Conformal Transformations with Multiple Scalar Fields},
  Phys.\ Rev.\ D~{\bf 81} (2010) 084044
  [arXiv:1003.1159 [gr-qc]].


\bibitem{Kaiser:2013sna}
  D.I.~Kaiser and E.I.~Sfakianakis,
 \emph{Multifield Inflation after Planck: The Case for Nonminimal Couplings,}
  Phys.\ Rev.\ Lett.~{\bf 112} (2014)  011302
  [arXiv:1304.0363 [astro-ph.CO]];\\
  K.~Schutz, E.I.~Sfakianakis and D.I.~Kaiser,
  \emph{Multifield Inflation after Planck: Isocurvature Modes from Nonminimal Couplings},
  Phys.\ Rev.\ D {\bf 89} (2014) 064044
  [arXiv:1310.8285 [astro-ph.CO]].

\bibitem{Gong:2016qmq}
  J.O.~Gong,
  \emph{Multi-field inflation and cosmological perturbations},
  Int.\ J.\ Mod.\ Phys.\ D {\bf 26}, no.~01, 1740003 (2017)
  [arXiv:1606.06971 [gr-qc]].


\bibitem{KLMattr}
R.~Kallosh and A.~Linde,
  \emph{Multi-field Conformal Cosmological Attractors},
  J. Cosmol. Astropart. Phys. {\bf 1312} (2013) 006
  [arXiv:1309.2015 [hep-th]];\\
 A.~Achucarro, R.~Kallosh, A.~Linde, D.~G.~Wang and Y.~Welling,
 \emph{Universality of multifield $\alpha$-attractors},
  J. Cosmol. Astropart. Phys. {\bf 1804} (2018) no.04,  028
  [arXiv:1711.09478 [hep-th]].
\bibitem{MSSM}
  M.N.~Dubinin, E.Yu.~Petrova, E.O.~Pozdeeva, M.V.~Sumin and S.Yu.~Vernov,
  \emph{MSSM-inspired multifield inflation},
  J. High Energy Phys. {\bf 1712} (2017) 036
  [arXiv:1705.09624 [hep-ph]];\\
M.N.~Dubinin, E.Yu.~Petrova, E.O.~Pozdeeva and S.Yu.~Vernov,
 \emph{MSSM inflation and cosmological attractors},
  Int.\ J.\ Geom.\ Meth.\ Mod.\ Phys.\  {\bf 15} (2018) 1840001
  [arXiv:1712.03072 [hep-ph]].

\bibitem{Gottlober:1993hp}
  S.~Gottlober, J.P.~Mucket and A.A.~Starobinsky,
  \emph{Confrontation of a double inflationary cosmological model with observations},
  Astrophys.\ J.\  {\bf 434} (1994) 417
  [arXiv:astro-ph/9309049].

\bibitem{delaCruz-Dombriz:2016bjj}
  A.~de la Cruz-Dombriz, E.~Elizalde, S.~D.~Odintsov and D.~Saez-Gomez,
  \emph{Spotting deviations from R$^2$ inflation,}
  J. Cosmol. Astropart. Phys. {\bf 1605} (2016) no.05,  060
  [arXiv:1603.05537 [gr-qc]].

\bibitem{Wang:2017fuy}
  Y.C.~Wang and T.~Wang,
  \textit{Primordial perturbations generated by Higgs field and $R^2$ operator},
  Phys.\ Rev.\ D {\bf 96} (2017) 123506
  [arXiv:1701.06636 [gr-qc]].

\bibitem{Ema}
Y.~Ema,
  \textit{Higgs Scalaron Mixed Inflation},
  Phys.\ Lett.\ B {\bf 770} (2017) 403
  [arXiv:1701.07665 [hep-ph]]
\bibitem{He:2018gyf}
   M.~He, A.~A.~Starobinsky and J.~Yokoyama,
  \emph{Inflation in the mixed Higgs-$R^2$ model,}
  J. Cosmol. Astropart. Phys. {\bf 1805} (2018) no.05,  064
  [arXiv:1804.00409 [astro-ph.CO]].

  \bibitem{Gorbunov:2018llf}
  D.~Gorbunov and A.~Tokareva,
 \textit{Scalaron the healer: removing the strong-coupling in the Higgs- and Higgs-dilaton inflations},
  Phys.\ Lett.\ B {\bf 788} (2019) 37
  [arXiv:1807.02392 [hep-ph]];\\
F.~Bezrukov, D.~Gorbunov, C.~Shepherd and A.~Tokareva,
  \emph{Some like it hot: $R^2$ heals Higgs inflation, but does not cool it},
  Phys. Lett. B \textbf{795} (2019) 657-665, [arXiv:1904.04737 [hep-ph]].
\bibitem{Damour1992} T. Damour and G. Esposito-Farese,
\textit{Tensor-multi-scalar theories of gravitation}, Class. Quantum Grav. {\bf 9} (1992) 2093-2176.

 \bibitem{Chervon:2013gm}
S.V.~Chervon, \textit{Chiral Cosmological Models: Dark Sector Fields Description}, Quantum Matter, \textbf{2} (2013) 71--82

  \bibitem{Chervon:2013nsm}
R.R. Abbyazov, S.V.~Chervon, V. Mueller
\textit{$\Lambda$CDM coupled to radiation: Dark energy and Universe acceleration},  Modern Physics Letters A. - World Scientific Publishing Co., v. 30, No. 26, 1550114 (11 pages), 2015

\bibitem{Nojiri:2005pu}
  S.~Nojiri and S.D.~Odintsov,
  \emph{Unifying phantom inflation with late-time acceleration: Scalar phantom-non-phantom transition model and generalized holographic dark energy},
  Gen.\ Rel.\ Grav.\  {\bf 38} (2006) 1285
  [arXiv:hep-th/0506212].

\bibitem{Capozziello:2018jya}
  S.~Capozziello, Ruchika and A.A.~Sen,
  \emph{Model independent constraints on dark energy evolution from low-redshift observations},
  Mon.\ Not.\ Roy.\ Astron.\ Soc.\  {\bf 484}, 4484 (2019)
  [arXiv:1806.03943 [astro-ph.CO]].


\bibitem{Vikman}
  A.~Vikman,
  \emph{Can dark energy evolve to the phantom?},
  Phys.\ Rev.\ D {\bf 71} (2005) 023515
  [arXiv:astro-ph/0407107].

\bibitem{QuntomREV}
  Y.F.~Cai, E.N.~Saridakis, M.R.~Setare and J.Q.~Xia,
  \emph{Quintom Cosmology: Theoretical implications and observations},
  Phys.\ Rept.\  {\bf 493} (2010) 1
  [arXiv:0909.2776 [hep-th]].


\bibitem{givanov81}
G.G.~Ivanov, \emph{Friedmann's cosmological models with a nonlinear scalar field}, Gravitation and Theory of Relativity, Kazan,
\emph{Kazan university publishing house} \textbf{18}  (1981) 54
%



\bibitem{chervon87}
 A.A.~Chaadaev and S.V.~Chervon. \textit{New class of cosmological solutions for a self-interacting scalar field }. Russian Phys. J., New York, v.56, No.7, pp. 725-730, 2013.
%

\bibitem{chervon102}
 I.V. Fomin and S.V.~Chervon. \textit{Exact and approximate solutions in the Friedmann cosmology }. Russian Phys. J., NY, v.60, issue 30, pp. 427-440, 2017.

\bibitem{Chervon:2017kgn}
S.V.~Chervon, I.V. Fomin and A. Beesham,
{\it The method of generating functions in exact scalar field cosmology},
 Eur.\ Phys.\ J.\ C {\bf 78} (2018) no.4,  301
  [arXiv:1704.08712 [gr-qc]].

\bibitem{chervon26}
S.V.~Chervon, \textit{Gravitational Field of the Early Universe I: Non-linear scalar field as the source}. Gravitation \& Cosmology \textbf{3}, No.2, p. 145-150 (1997) [arXiv:gr-qc/9706028].


\bibitem{Paliathanasis}
  A.~Paliathanasis, G.~Leon and S.~Pan,
 \emph{Exact Solutions in Chiral Cosmology},
  Gen.\ Rel.\ Grav.\  {\bf 51},  no.9,  106 (2019)
  [arXiv:1811.10038 [gr-qc]].;\\
    N.~Dimakis, A.~Paliathanasis, P.A.~Terzis and T.~Christodoulakis,
  \emph{Cosmological Solutions in Multiscalar Field Theory},
  Eur.\ Phys.\ J.\ C {\bf 79}, no. 7, 618 (2019)
  [arXiv:1904.09713 [gr-qc]].

\bibitem{ChervAbb}
%
R.R. Abbyazov, S.V.~Chervon, \textit{Unified Dark Matter and Dark Energy Description in a Chiral Cosmological Model }, Modern Physics Letters A. - World Scientific Publishing Co., v. 28, No. 8, 1350024 (19 pages), 2013;\\
R.R. Abbyazov, S.V.~Chervon, V. Mueller, \textit{$\sigma$CDM coupled to radiation: Dark energy and Universe acceleration}, Modern Physics Letters A. - World Scientific Publishing Co., v. 30, No. 26, 1550114 (11 pages), 2015.

\bibitem{EmU}
S.V.~Chervon, S.D. Maharaj, A. Beesham,  and A.S. Kubasov, \textit{Emergent Universe Supported by Chiral Cosmological Fields in 5D Einstein-Gauss-Bonnet Gravity }, Gravitation \& Cosmology, v.20, No. 3, pp.176-181, 2014;\\
A. Beesham,  S.V.~Chervon, S.D. Maharaj, A.S. Kubasov, \textit{Exact Inflationary Solutions Inspired by the Emergent Universe Scenario}, International Journal of Theoretical Physics, 54:884-895, 2015.

\bibitem{EGB}
S.D. Maharaj, A. Beesham, S.V.~Chervon, and A.S. Kubasov. \textit{New exact solutions for a chiral cosmological model in 5D EGB Gravity}, Gravitation and Cosmology, v. 23, No.4, pp. 375-380, 2017;\\
I.V. Fomin and S.V.~Chervon, \textit{Exact inflation in Einstein-Gauss-Bonnet Gravity}. Gravitation and Cosmology, v. 23, No.4, pp. 367-374, 2017;\\
%
I.V. Fomin and S.V.~Chervon, \textit{A new approach to exact solutions constructions in scalar cosmology with a Gauss-Bonnet term }. Modern Physics Letters A, v.32, No 30, p. 1750129, 2017.
%
\bibitem{Chervon:2018stfi}
S.V.~Chervon, A.S. Kubasov and K.A. Bolshakova, \textit{Cosmological inflation in tensor-multi-scalar theory of gravitation}, Space, Time and Fundamental Interactions, 2018, v.~1, pp. 67-81.

		\bibitem{AKV2} I.Ya.~Aref'eva, A.S.~Koshelev, and S.Yu~Vernov, \textit{Crossing the $w=-1$ barrier
		in the D3-brane dark energy model},  Phys. Rev. D \textbf{72} (2005)
		064017  [arXiv:astro-ph/0507067]

		\bibitem{Vernov06}  S.Yu~Vernov,
		\textit{Construction of Exact Solutions in Two-Field Models},
		Theor. Math. Phys. \textbf{155} (2008) 544--556 [arXiv:astro-ph/0612487]

		\bibitem{Arefeva:2009tkq}
		I.Ya.~Aref'eva, N.V.~Bulatov and S.Yu.~Vernov,
		\textit{Stable Exact Solutions in Cosmological Models with Two Scalar Fields},
		Theor.\ Math.\ Phys.\  {\bf 163} (2010) 788
		[arXiv:0911.5105 [hep-th]].


		\bibitem{Muslimov} A.G. Muslimov, \textit{On the Scalar Field Dynamics
		in a Spatially Flat Friedman Universe}, Class. Quant. Grav. \textbf{7}
		(1990) 231--237
		\bibitem{Townsend}
		K. Skenderis and P.K. Townsend, \textit{Hamilton--Jacobi method for
		Domain Walls and Cosmologies}, Phys. Rev. D \textbf{74} (2006) 125008,  [arXiv:hep-th/0609056];\\
	  P.K. Townsend, \textit{Hamilton-Jacobi Mechanics from
		Pseudo-Supersymmetry}, Class. Quant. Grav. \textbf{25} (2008) 045017,
		[arXiv:0710.5178]

		\bibitem{AKV} I.Ya.~Aref'eva,
		A.S.~Koshelev, and S.Yu.~Vernov, \textit{Exactly Solvable SFT Inspired
		Phantom Model}, Theor. Math. Phys. \textbf{148} (2006) 895--909, [arXiv:astro-ph/0412619]

\bibitem{Bazeia}
D. Bazeia, C.B. Gomes, L. Losano, and R. Menezes, \textit{First-order
		formalism and dark energy}, Phys. Lett. B \textbf{633} (2006) 415--419
		 [arXiv:astro-ph/0512197];\\
D. Bazeia, L. Losano, R. Rosenfeld,
\emph{First-order formalism for dust},
Eur. Phys. J. C \textbf{55} (2008) 113--117 [arXiv:astro-ph/0611770]


\bibitem{Andrianov:2007ua}
  A.A.~Andrianov, F.~Cannata, A.Yu.~Kamenshchik, and D.~Regoli,
\emph{Reconstruction of scalar potentials in two-field cosmological models},
J. Cosmol. Astropart. Phys. {\bf 0802} (2008) 015 [arXiv:0711.4300];\\
  M. R. Setare, J. Sadeghi,
\emph{First-order formalism for the quintom model of dark energy},
   	Int. J. Theor. Phys. \textbf{47} (2008) 3219--3225 [arXiv:0805.1117]


\bibitem{Rotova}
V.K. Shchigolev and M.P. Rotova,
\emph{Cosmological model of interacting tachyon field},
Mod. Phys. Lett. A \textbf{27} (2012) 1250086 [arXiv:1203.5030]

\bibitem{Harko:2013gha}
T.~Harko, F.S.N.~Lobo, and M.K.~Mak,
\emph{Arbitrary scalar field and quintessence cosmological models,}
  Eur.\ Phys.\ J.\ C {\bf 74} (2014) 2784
  [arXiv:1310.7167 [gr-qc]].



\bibitem{KTVV2013}
A.Yu.~Kamenshchik, A.~Tronconi, G.~Venturi, and S.Yu.~Vernov,
\textit{Reconstruction of Scalar Potentials in Modified Gravity Models},
Phys. Rev. D \textbf{87} (2013) 063503 [arXiv:1211.6272]


\bibitem{chervon20E}
S.V.~Chervon and V.M. Zhuravlev,
\textit{Exact solutions in cosmological inflationary models}.
Russ. Phys. J., New York, v.39, 1996, p.776-780.

\bibitem{chervon24}
S.V.~Chervon, \textit{ Self-interacting scalar field coupled to gravity: new exact solutions}.  Gravitation \& Cosmology, \textbf{5}, Suppl., 1999, p.9--14.
 (ISSN 0370-2693)

\bibitem{chervon25}
S.V.~Chervon, V.M. Zhuravlev and V.K. Shchigolev \textit{New exact solutions in standard inflationary models}. Phys. Lett. B \textbf{398}, p.269--273, 1997. [arXiv:gr-qc/9706031].
%

\bibitem{chervon35E}
S.V.~Chervon, V.M. Zhuravlev and V.K. Shchigolev. \textit{New classes of exact solutions in inflationary cosmology}. JETF, New York, v.~\textbf{87}, p.223--228, 1998.
%
\bibitem{chervon38E}
V.M. Zhuravlev and S.V.~Chervon. \textit{Cosmological Inflation Models Admitting Natural Emergence to the Radiation-Dominated Stage and the Matter Domination Era}. JETF, New York, v.\textbf{91}, p.227--238, 2000.
%
\bibitem{chervon41E}
S.V.~Chervon, V.M.~Zhuravlev. \textit{Comparative analysis of approximate and exact models in inflationary cosmology}. Russ. Phys. J., New York, v.\textbf{43}, No.1, p.11-17, 2000.



\bibitem{chervon79}
A.V.~Yurov, V.A.~Yurov, S.V.~Chervon and M.~Sami, \textit{Potential of total energy as superpotential in integrable cosmological models}, Theor. Math. Phys. \textbf{166} (2011) 259.
%

\bibitem{Barrow:2016qkh}
  J.D.~Barrow and A.~Paliathanasis,
  \textit{Observational Constraints on New Exact Inflationary Scalar-field Solutions},
  Phys.\ Rev.\ D {\bf 94}, no. 8, 083518 (2016)
  [arXiv:1609.01126 [gr-qc]].
%

 \bibitem{Chervon:2008zz}
 S.V.~Chervon and I.V.~Fomin,
 \textit{On calculation of the cosmological parameters in exact models of inflation},
  Grav.\ Cosmol.\  {\bf 14}, 163 (2008)
 [arXiv:1704.05378 [gr-qc]];

S.V.~Chervon, \textit{ Inflationary cosmology without restrictions on the scalar field potential},
General Relativity and Gravitation 36:1547-1553 (2004);

S.V.~Chervon, M. Novello and R. Triay, \textit{Exact cosmology and specification of an inflationary scenario }, Gravitation \& Cosmology, v.11, No.4,  p. 329--344, (2005).


\bibitem{Binetruy:2014zya}
		P.~Binetruy, E.~Kiritsis, J.~Mabillard, M.~Pieroni and C.~Rosset,
		\textit{Universality classes for models of inflation},
		J. Cosmol. Astropart. Phys.  {\bf 1504} (2015) 033 (arXiv:1407.0820)
  \bibitem{Pieroni:2015cma}
		M.~Pieroni,
		\emph{$\beta$-function formalism for inflationary models with a non minimal coupling with gravity},
		J. Cosmol. Astropart. Phys.   {\bf 1602} (2016)  012
		[arXiv:1510.03691 [hep-ph]].
		\bibitem{Binetruy:2016hna}
		P.~Binetruy, J.~Mabillard and M.~Pieroni,
		\emph{Universality in generalized models of inflation},
  J. Cosmol. Astropart. Phys. {\bf 1703} (2017) no.03,  060
  [arXiv:1611.07019 [gr-qc]].



\bibitem{DeWolfe}  O. DeWolfe, D.Z. Freedman, S.S. Gubser, A. Karch,
		{\it Modeling the fifth dimension with scalars and gravity},
		Phys. Rev.  D {\bf 62} (2000) 046008; [arXiv:hep-th/9909134]

\bibitem{MMSV}
 A.S. Mikhailov, Yu.S. Mikhailov, M.N. Smolyakov, I.P. Volobuev,
\textit{Constructing stabilized brane world models in
five-dimensional Brans-Dicke theory},
Class. Quantum Grav.
\textbf{24} (2007) 231--242 [arXiv:hep-th/0602143]; \\
 M.N.~Smolyakov and I.P.~Volobuev,
  \textit{Single-brane world with stabilized extra dimension},
  Int.\ J.\ Mod.\ Phys.\ A {\bf 23} (2008) 761 [arXiv:0705.4495]

\bibitem{Bazeia:2013dna}
  D.~Bazeia, A.~S.~Lobao, L.~Losano and R.~Menezes,
 \textit{First-order formalism for flat branes in generalized N-field models},
  Phys.\ Rev.\ D {\bf 88} (2013) 045001
  [arXiv:1306.2618 [hep-th]].

\bibitem{chervon89}
A.V. Astashenok, A.V. Yurov, S.V.~Chervon, E.V. Shabanov, M. Sami, \textit{New exact cosmologies on the brane}. Astrophys. Space Sci. \textbf{353}:319--328, 2014



\bibitem{Gursoy:2008za}
 U.~Gursoy, E.~Kiritsis, L.~Mazzanti, and F.~Nitti,
\textit{Holography and Thermodynamics of 5D Dilaton-gravity},
J. High Energy Phys. {\bf 0905} (2009) 033  [arXiv:0812.0792]
\bibitem{Aref'eva:2014sua}
  I.Ya.~Aref'eva, E.O.~Pozdeeva, and T.O.~Pozdeeva,
\textit{Holographic estimation of multiplicity and membranes collision in modified spaces ${AdS}_5$},
 	Theor. Math. Phys. \textbf{176} (2013) 861--872 [arXiv:1401.1180].


  \bibitem{deAlfaro1979}
V.~de Alfaro, S.~Fubini and G.~Furlan,
\textit{Gauge Theories and Strong Gravity}
Nuovo Cimento, v.\textbf{50}A, No.4, p. 523, 1979

\bibitem{Perelomov87}
A.M.~Perelomov, \emph{Chiral models: Geometrical aspects}, Phys. Rep. \textbf{146}, No.3, 135 (1987);\\
A.M.~Perelomov,
\textit{Solutions of the instanton type in chiral models}
Soviet Physics Uspekhi
(Uspekhi Fiz. Nauk), \textbf{134}, issue 4, pp. 577--609 (1981).

\bibitem{givanov1983}
G.G.~Ivanov, \emph{Symmetries, conservation laws, and exact solutions in nonlinear sigma models},
Theor. Math. Phys. \textbf{57} (1983) 981--987
[Teor.Mat.Fiz.,  \textbf{57}, No. 1, p.45, 1983]

\bibitem{Chervon1995}
S.V.~Chervon,
\textit{On the chiral model of cosmological inflation}.
Russ. Phys. J., New York, v.38, p. 539--543, 1995.

\bibitem{Novikov1990} S.P.~Novikov, A.T.~Fomenko,
\textit{Basic Elements of Differential Geometry and Topology}, Springer,
Series: Mathematics and its Applications, Vol. 60 (1990), ISBN 978-94-015-7895-0.

\bibitem{Maeda:1988ab}
  K.i.~Maeda,
  \emph{Towards the Einstein-Hilbert Action via Conformal Transformation},
  Phys.\ Rev.\ D {\bf 39} (1989) 3159.

\bibitem{Ruzmaikina}  T.V.~Ruzmaikina, A.A.~Ruzmaikin,
		\textit{Quadratic Corrections to the Lagrangian Density of the Gravitational Field and the Singularity},
 Sov. Phys. JETP \textbf{30} (1970) 372		
		
\bibitem{Skugoreva:2014gka}
M.M.~Ivanov, A.V.~Toporensky, \emph{Stable super-inflating cosmological solutions in f(R)-gravity},
 	Int. J. Mod. Phys. D \textbf{21}, No. 6 (2012) 1250051 [arXiv:1112.4194];\\
  M.A.~Skugoreva, A.V.~Toporensky and S.Yu.~Vernov,
  \emph{Global stability analysis for cosmological models with nonminimally coupled scalar fields},
  Phys.\ Rev.\ D {\bf 90} (2014) 064044 [arXiv:1404.6226]

\bibitem{Hoyle}
F.~Hoyle, G.~Burbidge and  J.V.~Narlikar,
\textit{A quasi-steady state cosmological model with creation of matter},
Astrophysical Journal. {\bf 410} (1993) no. 2, 437--457.

\bibitem{Banerjee}
S.K.~Banerjee and  J.V.~Narlikar,
\textit{The Quasi-Steady State Cosmology: A Problem of Stability},
Astrophysical Journal. {\bf 487} (1997) 69--72.

\bibitem{Sachs}
R~Sachs, J.V.~Narlikar, F.~Hoyle,
\textit{The quasi-steady state cosmology: analytical solutions of field equations and their relationship to observations},
Astronomy and Astrophysics. {\bf 313} (1996) 703--712.

\bibitem{Reis} A.G. Riess, S. Casertano, W. Yuan, L.M. Macri, D. Scolnic,
\emph{Large Magellanic Cloud Cepheid Standards Provide a $1\%$
Foundation for the Determination of the Hubble Constant and Stronger Evidence for Physics Beyond LambdaCDM},
Astrophys. J. \textbf{876} (2019) no.1, 85 	[arXiv:1903.07603 [astro-ph.CO]].

\bibitem{sami}
 P.~Singh, M.~Sami and N.~Dadhich,
  \emph{Cosmological dynamics of phantom field},
  Phys.\ Rev.\ D {\bf 68} (2003) 023522
  [arXiv:hep-th/0305110];\\
 M.~Sami and A.~Toporensky,
  \emph{Phantom field and the fate of universe},
  Mod.\ Phys.\ Lett.\ A {\bf 19} (2004) 1509
  [arXiv:gr-qc/0312009]

\bibitem{Nojiri:2003vn}
  S.~Nojiri and S.D.~Odintsov,
 \emph{Quantum de Sitter cosmology and phantom matter,}
  Phys.\ Lett.\ B {\bf 562} (2003) 147
  [arXiv:hep-th/0303117].


\bibitem{Nesseris:2006er}
  S.~Nesseris and L.~Perivolaropoulos,
  \emph{Crossing the Phantom Divide: Theoretical Implications and Observational Status,}
  J. Cosmol. Astropart. Phys. {\bf 0701}, 018 (2007)
  [arXiv:astro-ph/0610092].

\bibitem{Lazkoz}
R.~Lazkoz, G.~Le\'on, and I. Quiros, \textit{Quintom cosmologies with
arbitrary potentials}, Phys. Lett. B \textbf{649} (2007) 103--110
[arXiv:astro-ph/0701353];\\
R.~Lazkoz, G.~Le\'on, \textit{Quintom cosmologies admitting either
tracking or phantom attractors}, Phys. Lett. B \textbf{638} (2006)
303--309 [arXiv:astro-ph/0602590]
\bibitem{Setare}
M.R.~Setare, E.N.~Saridakis,\textit{ Quintom dark energy models with nearly
flat potentials}, Phys. Rev. D \textbf{79} (2009) 043005 [arXiv:0810.4775];\\
M.R.~Setare, E.N.~Saridakis, \textit{Quintom Cosmology with General
Potentials}, Int. J. Mod. Phys. D \textbf{18} (2009) 549--557 [arXiv:0807.3807]
 \bibitem{ENO}
 E.~Elizalde, Sh.~Nojiri, S.D.~Odintsov, \textit{Late-time cosmology in
(phantom) scalar-tensor theory: dark energy and the cosmic speed-up},
Phys. Rev. D \textbf{70} (2004) 043539 [arXiv:hep-th/0405034]

\bibitem{Fomin:2017xlx}
  I.V.~Fomin and S.V.~Chervon,
\textit{Exact and Approximate Solutions in the Friedmann Cosmology},
  Russ.\ Phys.\ J.\  {\bf 60} (2017) no.3,  427.




\end{thebibliography}
\end{document}